\input epsf

\documentclass{fic-l}

\newtheorem{theorem}{Theorem}[section]

\newtheorem{proposition}[theorem]{Proposition}
\newtheorem{conjecture}[theorem]{Conjecture}

\theoremstyle{definition}     
\newtheorem{definition}[theorem]{Definition}

\theoremstyle{remark}

\numberwithin{equation}{section}

\def\mylabel#1{\label{#1}   \rlap{\centerline{#1}}   }
\def\thlabel#1{\label{#1}   \proplabeL{#1} \hskip-3pt  }

\def\mylabel#1{\label{#1}}
\def\thlabel#1{\label{#1}}

\def\pd{\partial}
\def\refFtwoBCOV{4.2}

\begin{document}

\title{Counting BPS States via Holomorphic Anomaly Equations}

\author{Shinobu Hosono}
\address{Graduate School of Mathematical Sciences \\ 
University of Tokyo \\
Tokyo 153-8914, Japan}


\date{Jun, 2002}


\begin{abstract}
We study Gromov-Witten invariants of a rational 
elliptic surface using holomorphic anomaly equation in [HST1].  
Formulating invariance under the affine $E_8$ Weyl group symmetry, 
we determine conjectured  invariants, the number of BPS states,  
from Gromov-Witten invariants. We also connect our holomorphic 
anomaly equation to that found by Bershadsky,Cecotti,Ooguri and Vafa 
[BCOV1]. 
\end{abstract}

\maketitle

\section{Introduction and Main results}

Let $S$ be a surface obtained by blowing up nine base points 
of two generic cubics in $\bold P^2$. $S$ has an elliptic fibration 
$f: S \rightarrow \bold P^1$ and, in this note, we call it rational 
elliptic surface or $\frac{1}{2}$K3. 
(The latter name comes from the fact that 
$S$ has 12 singular fibers of Kodaira $I_1$ type while 
a generic elliptic K3 surface has 24.)

The surface $S$ is of considerable interest in the study of 
Gromov-Witten invariants and, in fact, has been providing a 
testing ground for (local) mirror symmetry [KMV] of Calabi-Yau threefolds 
and its applications to enumerative geometry. For example, in [HSS] 
the celebrated Modell-Weil group of $S$ has been connected to 
certain genus zero Gromov-Witten invariants of $S$. In [HST1], 
a certain recursion relation ({\it holomorphic 
anomaly equation}) was found, which determines the generating function 
of Gromov-Witten invariants of $S$ for all genera. The main purpose 
of this note is to present a detailed study of the solutions of the 
holomorphic anomaly equation. Also we study Gromov-Witten invariants 
using similar but more general holomorphic anomaly equation valid 
for all Calabi-Yau threefolds due to [BCOV1,2], and remark a nontrivial 
relation between two equations. Main results in this paper are 
{\bf Proposition 2.4}, {\bf Tables 2--5}, and {\bf Conjecture 4.3}. 

To describe the setting in more detail, let us consider a Calabi-Yau 
threefold $X$ which contain a rational elliptic surface $S$. 
Consider the moduli space of stable maps from genus $g$ curves with 
$n$ point on it to $S$. Then genus $g$ Gromov-Witten invariant 
$N_g(\beta)$ with $\beta \in H_2(S,\bold Z)$ is defined by 
\begin{equation}
N_g(\beta)=\int_{[\bar{\mathcal M}_{g,0}(S,\beta)]^{vert}} c(R^1\pi_*\mu^* 
N_{S/X})\;\;,
\mylabel{eqn:GWdef}
\end{equation}
where $N_{S/X}$ is the normal sheaf and $\mu: {\mathcal M}_{g,1}(S,\beta) 
\rightarrow S$ is the evaluation map and $\pi: {\mathcal M}_{g,1}(S,\beta) 
\rightarrow {\mathcal M}_{g,0}(S,\beta)$ is the forgetful map.

For some special $\beta$,  
using localization method of torus actions, we may calculate $N_g(\beta)$ 
directly based on the definition \eqref{eqn:GWdef}, see e.g. [Ko][KZ] 
for details. Another way to determine $N_g(\beta)$ is to use the 
calculational technique based on mirror symmetry conjecture in [CdOGP] 
and [BCOV1]. Although the latter way has great advantage in calculating 
Gromov-Witten invariants, its equivalence to the abstract definition 
(\ref{eqn:GWdef}) has been  
established in [G] and [LLY1] only for some restricted Calabi-Yau 
hypersurfaces, see also [CK] for more backgrounds. Our holomorphic anomaly 
equation for $S$ came from the calculational technique based on the mirror 
symmetry[HST1]. 

\vskip0.3cm

\noindent
(1) To reproduce the holomorphic anomaly equation more specifically, 
let $F$ and $\sigma$ in $H^2(S,\bold Z)$, respectively, be the fiber 
class and the class of a section of the elliptic fibration.  
Then consider the following summation over $\beta$; 
$$
N_g(d,n):=\sum_{\beta.\sigma=d,\; \beta.F=n} N_g(\beta)
$$
and define the corresponding generating function with formal variable $q$;
\begin{equation}
Z_{g;n}(q):=\sum_{d \geq 0} N_g(d,n)q^d  \;\;.
\mylabel{eqn:ZgnE}
\end{equation}
In [HST1], generalizing the result in [MNW] for $g=0$, 
it was found that:

\noindent
{\it (Holomorphic anomaly equation): The generating function $Z_{g;n}$ 
has the form 
\begin{equation}
Z_{g;n}(q)=P_{g,n}(E_2(q),E_4(q),E_6(q)) {q^{\frac{n}{2}} \over 
\eta(q)^{12n}} 
\mylabel{eqn:ZgnModular}
\end{equation}
with some quasi-modular form $P_{g,n} \in \bold Q[E_2,E_4,E_6]$ of weight 
$2g+6n-2$, where $E_2,E_4,E_6$ are Eisenstein series of weight two, four and 
six, respectively, and $\eta(q)=q^{\frac{1}{24}}\prod_{m>0}(1-q^m)$. 
Moreover $Z_{g;n}$ satisfies 
\begin{equation}
{\pd Z_{g;n} \over \pd E_2}={1\over 24}\sum_{g'+g''=g} \sum_{s=1}^{n-1} 
s(n-s) Z_{g';s}\;Z_{g'';n-s}+{n(n+1) \over 24} Z_{g-1;n} \;,
\mylabel{eqn:HAEa}
\end{equation}
with the initial data $Z_{0;1}(q)={q^{1\over2} E_4(q) \over \eta(q)^{12}}$. 
}

One of the interesting features of this equation is that, 
under certain additional vanishing conditions (gap condition) on 
$N_{g}(d,n)$, we can determine $Z_{g;n}(q)$ for 
all $g \geq 0$ and $n \geq 1$. Some explicit formulas are presented 
in the end of this section. 
In this paper, using the affine 
$E_8$ Weyl symmetry which arises as isomorphisms of rational elliptic  
surfaces [Lo][Do], 
we will determine $N_g(\beta)$ for $\beta \in H_2(S,\bold Z)$ with 
$(\beta,F)=n=1,2,3,4$ and 
$g=\frac{1}{2}\{ (\beta,\beta)-(\beta,F)+2 \}  \leq 10$. 
(Proposition 2.4 and Tables 2--5.)

\vskip0.3cm

\noindent 
(2) Another important aspect of Gromov-Witten invariants is that 
the invariants take values in $\bold Q$, however these can be related  
to {\it integer} ``invariants'' which, for example, may be identified with 
the number of (rational) curves in Calabi-Yau manifold. The relation to 
the integer ``invariants'' has appeared as multiple cover formula in [CdOGP] 
and [AM] for genus $g=0$, and its most general form has been proposed by 
Gopakumar and Vafa giving physical meanings for the integer ``invariants'', 
i.e. the number of {\it BPS states}: 

\noindent
{\it (Gopakumar-Vafa conjecture): Gromov-Witten invariants $N_g(\beta)$ 
are related to integer invariants $n_g(\beta)$ (the number of BPS states 
of charge $\beta$) by 
\begin{equation}
N_g(\beta)=\sum_{k | \beta} \sum_{h=0}^g C(h,g-h) k^{2g-3} n_h(\beta/k) \;,
\mylabel{eqn:GVformula}
\end{equation}
where $\sum_{k | \beta}$ means the summation over positive integer $k$ 
which divide the integral class $\beta$, and $C(h,g-h)$ is the rational 
number defined by 
$$
\left( {\sin(t/2) \over t/2 } \right)^{2g-2}
=\sum_{h=0}^\infty C(g,h) t^{2h} \;\;.
$$
}

Our result in this respect is that we verify the integrality of $n_g(\beta)$ 
up to $g \leq 10$ and $\beta.F \leq 4$ for rational elliptic surface $S$. 
(Tables 2--5.) 
Gopakumar and Vafa have also proposed 
that the integer ``invariants'' $n_g(\beta)$ should be geometric 
invariants on the moduli space of D2 branes of charge $\beta$, i.e. 
suitable moduli space of curves of a fixed homology class $\beta$ and 
with local system on it. 
Precise mathematical definition of the 
moduli space of D2 branes $\mathcal M_\beta(X)$ 
has been proposed in [HST2] for Calabi-Yau threefold $X$ 
with an ample class $L$. There the moduli space 
$\mathcal M_\beta(X)$ is defined as the normalization 
of the moduli space of semistable sheaves of pure 
dimension one with its support having homology class $\beta$, and also 
with a fixed Hilbert polynomial 
$P(m)=d m+1$ $(d=L\cdot \beta)$. Some numbers $n_g(\beta)$ 
have been explained from this definition[HST2]. 
We will provide a brief scketch in sect.3.3 about 
the expected geometrical interpretation about the numbers 
$n_g(\beta)$, although its detailed study is beyond 
the scope of our note.  Here we remark that in  case 
of elliptic surfaces, like $\frac{1}{2}$K3, the moduli 
spaces of D2 branes may be mapped to the moduli space 
of stable sheaves on the surface under fiberwise 
Fourier-Mukai transformations, see for example [MNVW], 
[Yo],[HST3].

\vskip0.3cm
\noindent
(3) The most general form of the holomorphic anomaly equation which is 
applicable, in principle, to arbitrary Calabi-Yau threefold is known 
in [BCOV1,2]. We will connect our holomorphic anomaly equation 
(\ref{eqn:HAEa}) to a certain limit (local mirror symmetry limit) 
of the equations in [BCOV1,2]. 
We will make explicit comparisons of these two equations for $g=2,3$, 
and conjecture their equivalence. (Conjecture 4.3). Also we will 
find a nontrivial relation in the holomorphic ambiguities of   
these equations.

\vskip0.5cm
Finally, for reader's convenience, we present here some explicit forms of   
solutions of the holomorphic anomaly equation (\ref{eqn:HAEa}):   
\begin{eqnarray*}
&&
 Z_{1,1}(q)=\frac{E_2(q) E_4(q)}{\prod_{n\geq1}(1-q^n)^{12}}  \;,\;\;
 Z_{2,1}(q)=\frac{E_4(q) ( 5 E_2(q)^2 + E_4(q) )}
            {1440 \prod_{n\geq1}(1-q^n)^{12}}   \\
&&
Z_{3,1}(q)=\frac{E_4(q)( 35 E_2(q)^3 + 21 E_2(q) E_4(q) + 4 E_6(q) )}
{362880 \prod_{n\geq1}(1-q^n)^{12}}  \;\;, 
\end{eqnarray*}

\begin{eqnarray*}
& Z_{0,2}(q)=&\frac{  E_2(q) E_4(q)^2 + 2 E_4(q) E_6(q)  }
                   { \prod_{n\geq1}(1-q^n)^{24} }  \;, \\  
& Z_{1,2}(q)=&\frac{  10 E_2(q)^2 E_4(q)^2 + 9 E_4(q)^3 + 
                      24 E_2(q) E_4(q) E_6(q) + 5 E_6(q)^2  }
               { 1152  \prod_{n\geq1}(1-q^n)^{24} }
                \;\;, \\
\end{eqnarray*}

\begin{eqnarray*}
& Z_{2,2}(q) =& \big(   
                190 E_2(q)^3 E_4(q)^2 + 417 E_2(q) E_4(q)^3 + 
                540 E_2(q)^2 E_4(q) E_6(q) + \\   
&             &  356 E_ 4(q)^2 E_6(q) + 
                        225 E_2(q) E_6(q)^2  \big) 
                \frac{1} {207360 \prod_{n\geq1}(1-q^n)^{24} } 
             \;, \\
& Z_{3,2}(q)=& \big( 2275 E_2(q)^4 E_4(q)^2 + 8925 E_2(q)^2 E_4(q)^3 + 
                      3540 E_4(q)^4 +  \\
&              &7560 E_2(q)^3 E_4(q) E_6(q)  
                       + 14984 E_2(q) E_4(q)^2 E_6(q) +  \\
&              &4725 E_2(q)^2 E_6(q)^2 + 4071 E_4(q) E_6(q)^2  \big)
               \frac{1}{ 34836480 \prod_{n\geq1}(1-q^n)^{24} } \;\;.
\end{eqnarray*}  

\noindent 
For $n=1$ a closed formula valid for all genus is known in [HST1]. 

\vskip0.3cm
\noindent
{\bf Acknowledgements:} 
The author would like to thank Department of Mathematics, 
Harvard University for the hospitality and support during 
his visit in the summer 1999 and 2001-2002. Most of this work has been 
done during his visit in the summer, 1999.  He would like to thank 
B. Andreas, R. Donagi, A. Klemm, B. Lian,  C-H Liu, K.Oguiso, 
M.-H.Saito, A.Takahashi, C. Vafa, and S.-T. Yau for their valuable 
discussions and comments.  Especially  he would like to thank 
M.-H.Saito, A.Takahashi and R. Donagi for 
their collaboration at early stage of this work. This research 
is also supported in part by Education 
Ministry of Japan.


\section{Generating function and affine $E_8$ Weyl orbits}

\subsection{Notations}

Let $S$ be a rational elliptic surface, i.e. $\bold P^2$ blown up 
at nine base points of two generic cubics. We denote by $e_i$ the 
cohomology class of exceptional curve $D_i\; (i=1,\cdots,9)$. 
Let $H$ be the pullback of the class of a line in $\bold P^2$. 
The second cohomology $H^2(S,\bold Z)$ is generated by $H, e_1,\cdots, e_9$; 
$$
H^2(S,\bold Z)=\bold Z H \oplus \bold Z e_1 \oplus \cdots \oplus \bold Z e_9
\;\;.
$$
Due to Poincar\'e duality, $H^2(S,\bold Z)$ becomes unimodular lattice 
with respect to the natural intersection pairing (cup product) 
$(*,**): H^2(S,\bold Z) \times H^2(S, \bold Z) \rightarrow \bold Z$. 
$S$ has an elliptic fibration $f: S \rightarrow \bold P^1$ with the class of 
the fiber given by 
$$
F=3H-e_1-e_2-\cdots -e_9 \;\;.
$$ 
In this note we fix an exceptional curve $D_9$ as the zero section. 
Then it is known that the orthogonal lattice,  
$$
\langle e_9, F \rangle^\perp:=\{ x \in H^2(S,\bold Z) \; | \; 
(x,e_9)=(x,F)=0 \;\} 
$$
is isomorphic to the lattice $E_8(-1)$, i.e. the $E_8$ root lattice with 
its pairing multiplied by $-1$.

\subsection{Root system}

Let $V$ be a real vector space and $V^*$ be its dual. A finite set 
$B$ of linearly independent vectors in $V$ together with an injection 
$\vee: B \rightarrow V^*, \alpha \rightarrow \alpha^\vee$ is called 
{\it root basis} if the following conditions are satisfied: 
(i) $B^\vee=\{ \alpha^\vee | \alpha \in B \}$ are linearly independent, 
(ii) $\alpha^\vee(\alpha) =-2$ for all $\alpha$, 
(iii) $\beta^\vee(\alpha), \alpha\not=\beta,$ are nonnegative integers, 
(iv) $\beta^\vee(\alpha) =0$ implies $\alpha^\vee(\beta) =0$. 
A root basis is called {\it symmetric} if $\alpha^\vee(\beta)=
\beta^\vee(\alpha)$ holds.

When $V$ is equipped with a non-degenerate pairing $(\;,\;):V \times V 
\rightarrow \bold R$ and we define $\vee: B \rightarrow V^*$ by 
\begin{equation}
\alpha^\vee(x)=(\alpha,x) \;\;, \;\; (x \in V), 
\mylabel{eqn:veedef}
\end{equation}
then the first property (i) is easily verified, and also 
$\alpha^\vee(\beta)=\beta^\vee(\alpha)$.  We will soon 
restrict our attention to a root basis $B$ in 
$V=H^2(S,\bold Z)\otimes \bold R$ with the  
injection $\vee$ defined by the nondegenerate cup product. 

Let $(B, V)$ be a symmetric root basis and write $B=\{ \alpha_0,\alpha_1, 
\cdots, \alpha_r \}$. The (symmetric) matrix 
\begin{equation}
A:=\left( a_{ij} \right)=
\left(  \alpha^\vee_i(\alpha_j) \right)_{0\leq i,j \leq r} 
\mylabel{eqn:cartan}
\end{equation}
is called the {\it Cartan matrix} of $B$. We may define 
a lattice structure on the group $Q=\bold Z \alpha_0 + \bold Z \alpha_1 + 
\cdots \bold Z \alpha_r$ by setting the bilinear form $(\alpha_i,\alpha_j)_Q 
=a_{ij}$. This is called the {\it root lattice} of $B$. 
Note that when $\vee$ is defined by (\ref{eqn:veedef}), the bilinear form 
$(\,,\,)_Q$ on the root lattice coincides with the pairing $(\;,\;)$ on 
$V$ restricted to $Q \subset V$. 
(However, it should be noted that the restriction 
of the nondegenerate pairing to $Q$ is not necessarily nondegenerate 
on $Q$.)  For any $\alpha_i \in B$, we define 
a {\it fundamental reflection} by 
\begin{equation}
s_i(x):=s_{\alpha_i}(x)=x+ \alpha^\vee_i(x) \alpha_i 
\quad (x \in V).
\mylabel{eqn:sa}
\end{equation}
Since $a_{ij}=\alpha_i^\vee(\alpha_j)=a_{ji}$, one may verify that 
$s_i$ is an element of the orthogonal group $O(Q)$ of the root lattice $Q$. 
The {\it Weyl group} of $B$ is a discrete subgroup of $O(Q)$ 
which is generated by fundamental reflections. The {\it fundamental 
Weyl chamber} $C$ is defined by 
$$
C=\{ x \in V \vert \alpha^\vee(x)  >0 \; (\alpha \in B) \},
$$
and $w(C)$ for some $w \in W$ is called simply a {\it chamber}. 
For each subset $Z \subset B$, we define a {\it fundamental facet} by 
$$
Facet_Z:=\{ x \in V \vert \alpha^\vee(x) =0 \; \text{for} \; 
\alpha \in Z \;\text{and } \; \alpha^\vee(x) >0 \; 
\text{for} \; \alpha \in B \setminus Z \}.
$$
Note that $Facet_\phi=C$ and the closure $\bar C$ of $C$ is the 
disjoint union of the fundamental facets. The $W$-orbit of $\bar C$ 
is called the {\it Tits cone} and denoted $I:=\bigcup_{w \in W} w (\bar C)$. 
Tits cone is a convex cone in $V$. It is known that the Weyl group 
acts properly discontinuously on the interior $\overset{\;_\circ} I$ 
of $I$ and $\bar C$ 
is a fundamental domain for this action. Also it is known that 
the Weyl group acts simply and transitively on the set of chambers, 
$\{ w(C) | w \in W \}$. 

The elements $\Lambda_j$ in $V$ satisfying $\alpha^\vee_i(\Lambda_j)
=\delta_{ij}$ are called {\it fundamental weights}. Note that 
fundamental weights are determined up to an elements $F_B$.

\subsection{Root system defined in $H^2(S,\bold Z)$} Here we introduce 
a root basis in $V=H^2(S,\bold Z)\otimes \bold R$ following [Lo]. 
Let us define $\alpha_0=e_8-e_9, \alpha_i=e_i-e_{i+1} \; (1\leq i \leq 7)$ 
and $\alpha_8=H-e_1-e_2-e_3$ and consider a finite set in $V$ 
$$
B=\{ \alpha_0,\alpha_1,\cdots,\alpha_8 \} \;\;.
$$
Since the cup product on $H^2(S,\bold Z)$ is nondegenerate, so is 
its scalar extension to $V$. By this nondegenerate form and 
(\ref{eqn:veedef}), we define the injective map $\vee: B \rightarrow V^*$. 
Then it is easy to verify that $(B,V)$ is in fact a root basis defined 
in 2.2, and also that the Cartan matrix of $B$ defined by (\ref{eqn:cartan}) 
coincides 
with that of the affine $\hat E_8(-1)$[Kac]. (In fact, the root basis 
is of {\it affine type}, which characterized by the properties: 
(i) it is irreducible, (ii) the Cartan matrix is of corank one 
and (iii) $W_X:=\langle s_{\alpha} | \alpha \in X \rangle$ is finite 
group for any proper subset $X \subset B$. See [Kac] for more details.) 
The Weyl group associated to 
this root basis is called {\it affine Weyl group} of $E_8(-1)$, and 
will be denoted by $W_{\hat E_8}$. By definition, the root lattice 
$(Q, (\;,\;)_Q)$ is naturally a sublattice of $(H^2(S,\bold Z),(\;,\;))$, 
and we may verify directly that  
$$
H^2(S,\bold Z)=Q \oplus \bold Z F = \bold Z \alpha_0 \oplus \bold Z \alpha_1 
\oplus \cdots \bold Z \alpha_8 \oplus \bold Z F , 
$$
as a lattice. Also we verify $Facet_B=\bold R \; F$. The affine Weyl group 
is an subgroup of $O(Q)$, and also may be regarded as a subgroup of 
$O(H^2(S,\bold Z))$ since it acts trivially on $\bold Z \,F$.

The Tits cone $I$ is known in [Lo, Proposition (3.9)] to be the union of 
the half space $\{ x \in V | (x,F)>0 \}$ and the facet $Facet_B=\bold R \; F$. 

The fundamental weights $\Lambda_i \in V$ (, s.t. $\alpha^\vee_i(\Lambda_j)=
\delta_{ij}$, ) are determined up to $Facet_B$. 
Since the lattice $H^2(S,\bold Z)$ 
is unimodular, we may take $\Lambda_i$ in $H^2(S,\bold Z)$ up to 
$\bold Z \, F$.  Fixing this ambiguity by hand, we define 
\begin{equation}
\begin{matrix}
&\Lambda_0=e_9,\;\;
\Lambda_1=H-e_1,\;\;
\Lambda_2=2H-e_1-e_2,\;\; 
\Lambda_3=3H-e_1-e_2-e_3,  \\
&
\Lambda_4=3H-e_1-e_2-e_3-e_4\;\;,\;\;
\Lambda_5=F+e_6+e_7+e_8+e_9 \;\;,\;\;  \\
&
\Lambda_6=F+e_7+e_8+e_9 \;\;,\;\; 
\Lambda_7=F+e_8+e_9 \;\;,\;\;
\Lambda_8=H \;.\hfil 
\end{matrix}
\mylabel{eqn:Fweight}
\end{equation}

\noindent
{\bf Remark.} 
(1)We note that the zero-th root may be written by $\alpha_0 = F -\theta$, 
where $\theta=2\alpha_1 + 4 \alpha_2 + 6\alpha_3 + 5\alpha_4 + 4\alpha_6 
+ 2\alpha_7 + 3\alpha_8$ is the {\it highest root} of the (classical) 
root basis $B^{cl}:=\{ \alpha_1,\cdots,\alpha_8 \}$.  

(2) We may extend linearly the injective map on a (symmetric) root basis  
$\vee: B \rightarrow V^*$  to the root lattice $\vee: Q \rightarrow 
V^*$, $\sum_k m_k \alpha_k \mapsto \sum_k m_k \alpha_k^\vee$. Then 
the simple reflection $s_{\alpha_i}$ defined for $\alpha_i \in B$ 
by (\ref{eqn:sa}) may be extended to $r_\alpha$ for $\alpha \in Q$ with 
$\alpha^\vee(\alpha)=-2$. The highest root $\theta$ is a so-called 
{\it real root}, i.e. a root $\alpha$ such that $\alpha=w(\alpha_i)$ 
for some $w \in W$ and $\alpha_i \in B$. From this, we see 
$\theta^\vee(\theta)=(\theta,\theta)=-2$ and also $r_\theta \in W$ since 
we have the relation  
$r_{\alpha_i}\circ r_{\alpha_j}\circ r_{\alpha_i}=r_{r_{\alpha_i}(\alpha_j)}$. 
Now we define {\it translation} $t_\gamma: Q \rightarrow Q \; (\gamma \in 
E_8(-1))$ by  
\begin{equation}
t_{\gamma}(\beta) = \beta+(F,\beta)\gamma - 
\{\frac{1}{2}(F,\beta)(\gamma,\gamma) +(\beta,\gamma)\} F \;\;, 
\mylabel{eqn:translation}
\end{equation}
which satisfy $t_\gamma\circ t_{\gamma'}=t_{\gamma+\gamma'}$,  
and consider a group of translations 
$T:=\{ t_\gamma | \; \gamma \in E_8(-1) \}$. Then 
we may verify the following relations; 
$$
r_{\alpha_0}\circ r_\theta = t_{-\theta} \;\;,\;\; 
r_\alpha \circ t_{-\theta} \circ r_\alpha = t_{-r_\alpha(\theta)} .
$$
In fact, it is known (see e.g. [Kac]) that the affine Weyl group 
$W_{\hat E_8}$ is a semi-direct product of the translation group $T$ and 
the {\it classical} Weyl group $W_{E_8}:=\langle r_{\alpha_1},\cdots, 
r_{\alpha_8} \rangle$; 
\begin{equation}
W_{\hat E_8}=W_{E_8} |\hskip-5pt\times T \;\;.
\mylabel{eqn:semiDP}
\end{equation}

\subsection{$Z_{g;n}$ and orbit decompositions} 

Let $s_{\alpha_i} (0\leq i \leq 8)$ be  reflections defined 
in (\ref{eqn:sa}), and consider their actions on the cohomology 
basis $s_{\alpha_i}: 
H,e_1,\cdots,e_9$ $\mapsto s_{\alpha_i}(H),s_{\alpha_i}(e_1),\cdots,
s_{\alpha_i}(e_9)$. For $i=0,\cdots,7$, the actions are 
simply interchanges $e_j \leftrightarrow e_{j+1}$. For $s_{\alpha_8}$, 
we have 
\begin{eqnarray*}
&s_{\alpha_8}(H)=2H-e_1-e_2-e_3 \;\;,\;\;
s_{\alpha_8}(e_1)=H-e_2-e_3 \;\;,\;\; \\
&s_{\alpha_8}(e_2)=H-e_1-e_3 \;\;,\;\;
s_{\alpha_8}(e_3)=H-e_1-e_2 \;\;,
\end{eqnarray*}
and $s_{\alpha_8}(e_k)=e_k \; (4\leq k \leq 9)$. 
Here we see, for example, that the class $s_{\alpha_8}(e_1)$ represents  
that of the line passing through the points $p_2$ and $p_3$ where we 
blow up in $\bold P^2$ to obtain $S$. Each class represents a smooth 
rational curve with self-intersection -1, which can be contracted. Therefore 
for each $s_{\alpha_i} \; (0 \leq i \leq 8)$, the classes 
$s_{\alpha_i}(e_1), \cdots, s_{\alpha_i}(e_9)$ represent 
the -1 curves which we can contract. Contracting these 9 curves to 
points $p_1', \cdots, p_9'$, we obtain $'\bold P^2$ which is birational 
to the original $\bold P^2$ used to define $S$. From this viewpoint, 
we may regard the class $s_{\alpha_i}(H)=:H'$ as the pullback of the 
class of a line in $'\bold P^2$, and $s_{\alpha_i}(e_k)=:e_k'$ as 
the class of the exceptional divisor for the blowing up at $p_k'$. 
The configuration of $p_1',\cdots,p_9'$ in $'\bold P^2$ differs 
from that of $p_1,\cdots,p_9$ in $\bold P^2$, and thus blowing 
up these points results in rational elliptic surface $S'$ with 
different complex structure from $S$. However by construction, 
$S'$ is identical to $S$. That is, there is an isomorphism 
between the two rational elliptic surfaces $S$ and $S'$ with 
different complex structures. (See [Lo, Theorem (5.3)] for 
Torelli type theorem for rational surfaces.)

Now we may combine this isomorphism with the invariance of 
Gromov-Witten invariants under the deformations. To describe it 
precisely, let us write $N_g^S(\beta) \; (\beta \in H^2(S,\bold Z))$ 
the Gromov-Witten invariants for the surface $S$, and similarly 
$N_g^{S'}(\beta') \; (\beta' \in H^2(S',\bold Z))$ for the surface 
$S'$. For example, let us assume $S'$ is defined as above for the 
reflection $s_{\alpha_8}$ and use the notations for $H,e_1,\cdots,e_9$ 
and $H',e_1',\cdots,e_9'$ introduced above. Then we have, for example;
$$
N_g^S(e_1)=N_g^{S'}(e_1')=N_g^{S}(H-e_2-e_3) \;\;,
$$
where the first equality is the invariance under the deformations and 
the second follows the isomorphism $\Phi: S \cong S'$. In the exactly 
same way, we have the equality $N_g^S(\beta)=N_g^S(s_{\alpha_i}(\beta))$ 
for all reflections $s_{\alpha_i} \; (i=0,1,\cdots,8)$. Since the affine 
Weyl group $W_{\hat E_8}$ is generated by the reflections $s_{\alpha_i}$, 
we have:

\begin{proposition} \thlabel{th:weylNg} 
$$
N_g(\beta)=N_g(\omega(\beta)) \;\;,\;\; (\beta \in H^2(S,\bold Z), 
\omega \in W_{\hat E_8})
$$
\end{proposition}

In what follows we will utilize this invariance to study the 
(solutions of the) holomorphic anomaly equation (\ref{eqn:HAEa}).  
As a result, in the next section, we will determine the numbers 
$N_g(\beta)$ for several $\beta \in H^2(S,\bold Z)$.  
The idea is simply to make the orbit decomposition of the generating  
function: 

\begin{definition} \thlabel{def:ZgnB} 
We define the {\it character of the generating function (or simply 
generating function}), ${\mathcal Z}_{g;n}: H^2(S,\bold Z)\otimes \bold C 
\rightarrow \bold C^*$ by  
\begin{equation}
\mathcal Z_{g;n}:= \sum_{\beta \in H^2(S,\bold Z), (\beta,F)=n} 
N_{g}(\beta) e^{2\pi \sqrt{-1} \beta}
\;\; (n>0)
\mylabel{eqn:characterZ}
\end{equation}
where $e^{2\pi \sqrt{-1} \beta}$ is the character defined 
by $e^{2\pi \sqrt{-1} \beta}(c):=e^{2\pi \sqrt{-1} (\beta,c)}$ 
for $c \in H^2(S,\bold Z)\otimes \bold C$
with the cup product $(\;,\;)$ extended to over $\bold C$.
\end{definition}

{\bf Remark} (1) The condition $(\beta,F)=n$ restricts the classes $\beta$ 
to those of $n$-sections. Since this condition is obviously invariant 
under the Weyl group action, we define  $\mathcal Z_{g;n}$  
restricting the sum over $\beta$'s of $n$-sections. 

(2) The generating function $Z_{g;n}(q) \; (q=e^{2\pi \sqrt{-1} \tau})$ 
introduced in (\ref{eqn:ZgnE}) is the character $\mathcal Z_{g;n}$ 
evaluated by $\tau \sigma$ with a class of (positive) section 
$\sigma=e_9+F$, i.e.,
$
Z_{g;n}(q)=\mathcal Z_{g;n}(\tau \sigma) \;\;.
$

\vskip0.5cm

By the general theory of Gromov-Witten 
invariants[KM], to have non-vanishing Gromov-Witten invariants 
$N_g(\beta)$ it is necessary that  $\beta$ represents a class of 
effective and connected (but not necessarily irreducible) 
divisor.  For connected and effective divisor class $\beta$, 
we have $(\beta,F)\geq0$ and the equality holds only if $\beta=k F$ 
for some positive integer $k$.  If we omit these rather trivial cases 
$\beta=k F$ from our consideration, we see that the condition 
$(\beta,F)>0$ coincides with that $\beta$ belongs to an integral 
class contained in the Tits cone. Now it is obvious from Proposition 
\ref{th:weylNg} that the invariant $N_g(\beta)$ is determined by 
its value for $\beta$ in the closure $\bar C$ of the fundamental Weyl 
chamber. 

The integral elements $\lambda$ in the 
fundamental Weyl chamber are called {\it dominant weight 
of level} $n(>0)$ if they satisfies $(\lambda, F)=n$. If $\lambda$ is 
dominant integral weight of level $n$, then so is $\lambda + a F$ for 
arbitrary integer $a$. To choose this $a$ as small as possible, we impose 
the following numerical conditions;
$$
\begin{matrix}
& (1) & \text{the arithmetic genus} \hfill \cr 
& &
g_{\lambda'}=
\frac{1}{2}\{(\lambda',\lambda')+2-(\lambda',F) \} \geq 0 \;\;, \cr
& & \text{and } g_{\lambda'}  \text{is minimum.} \hfill \cr
& (2) & \text{ if $n\geq2$ then $d\geq1$ and $a_1,\cdots,a_9 \geq 0$ for  
$\lambda'=d H -a_1e_1 - \cdots -a_9 e_9$. }
\end{matrix}
$$

We will call the dominant weights satisfying (1) and (2) {\it minimal}.

\begin{definition} \thlabel{def:PP} We denote  
the set of minimal dominant weights of level $n$ by $\mathcal P_{+,n}^{min}$, 
i.e. 
$
\mathcal P_{+,n}^{min}:=\{ \lambda \in H^2(S,\bold Z) \;|\; 
(\lambda, \alpha_i)\geq 0 \;(i=1,\cdots,8), 
(\lambda, F)=n, \lambda \text{: minimal}  \} 
\;\;.
$
\end{definition}

It is easy to verify that each fundamental weight  
$\Lambda_i$ introduced in (\ref{eqn:Fweight}) is minimal 
as well as dominant.  Note that addition of minimal dominant weights 
results in a dominant weight, however the minimality of weights is not 
preserved.  Now it will be convenient to define the addition among 
the minimal dominant weight by 
\begin{equation}
\lambda + \lambda' := \text{ minimal dominant weight in } \lambda+\lambda' + 
\bold Z F ,
\mylabel{eqn:addition}
\end{equation}
for minimal dominant weights $\lambda, \lambda'$. Hereafter we write 
the fundamental weights $\Lambda_0,\Lambda_1 \cdots,\Lambda_8$ by 
$\lambda_0, \lambda_1,\cdots,\lambda_8$ with this understanding for 
the addition. In Table 1, elements in $\mathcal P_{+,n}^{min}$ are 
listed for $n\leq 4$.

Now we are ready to accomplish the orbit decomposition of the character 
(\ref{eqn:characterZ}):

\begin{proposition} \thlabel{th:orbit} The character $\mathcal Z_{g;n}$ 
is decomposed into the orbits by 
\begin{equation}
\mathcal Z_{g;n}=\sum_{\lambda \in \mathcal P_{+,n}^{min}} 
\mathcal Z_{g,\lambda} \; P_{\lambda} \;\;,
\end{equation}
where 
\begin{equation}
\mathcal Z_{g,\lambda}:=
\sum_{a\in \bold Z} N_g(\lambda+a F) e^{2\pi \sqrt{-1}(\lambda+aF)}
\;\;,\;\;
P_{\lambda}:=\sum_{\omega \in W_{\hat E_8}(\lambda)} 
e^{2\pi \sqrt{-1}(\omega(\lambda)-\lambda)} \;\;,
\mylabel{eqn:ZandP}
\end{equation}
with $W_{\hat E_8}(\lambda):=W_{\hat E_8}/(\text{stabilizer of }\lambda)$. 
\end{proposition}

\begin{proof}
Since the integral classes $\beta$ with $(\beta,F)=n >0$ are contained 
in the Tits cone, for each Weyl orbit we may take a unique representative 
in the closure $\bar C$ of the fundamental Weyl chamber. Then we have 
\begin{align*}
\mathcal Z_{g;n}&=\sum_{\beta \in H^2(S,\bold Z), (\beta,F)=n} 
N_g(\beta) e^{2\pi \sqrt{-1} \beta} \\
&=\sum_{\lambda \in \mathcal P^{\min}_{+,n}} \; \sum_{a \in \bold Z} \; 
  \sum_{\omega \in W_{\hat E_8}(\lambda)} 
        N_g(\omega(\lambda)+ aF ) 
       e^{2\pi \sqrt{-1}( \omega(\lambda) + a F) } \\
&=\sum_{\lambda \in \mathcal P_{+,n}^{min}} 
  \left(\sum_{a\in \bold Z} N_g(\lambda+ a F) 
       e^{2\pi \sqrt{-1}(\lambda+a F)}\right) 
\left(\sum_{ w \in  W_{E_8}(\lambda)} 
e^{2\pi \sqrt{-1} (\omega (\tilde\lambda)-\lambda)}
\right), 
\end{align*}
where we remark that if $\lambda$ sits in the walls of $\bar C$, it  
has nontrivial stabilizers. Also the summation over $a$ has in fact 
lower bound(, see Remark below).  
\end{proof}

{\bf Remark} By general property of Gromov-Witten invariants, 
we have $N_g(\lambda + a F)=0$ unless $\lambda+a\,F$ is effective. 
Since $\lambda+a\,F$ is not effective for $a<<0$, we have a lower 
bound $a_0$ for the summation over $a\in \bold Z$ in the above proposition. 
For the examples, which are listed in this paper (Table 2--5), the 
lower bounds turn out in fact to be zero, i.e. $a_0=0$.  
The invariance under the affine Weyl group was used implicitly [MNVW] 
in making orbit decompositions and also discussed in general in a recent 
paper [Iq] which is similar to ours.

\vskip0.5cm

The character $P_\lambda \;(\lambda \in \mathcal P_{+,n}^{min})$ 
represents a summation over the Weyl orbit which is parametrized 
by $\lambda + a F \; (a\geq 0)$. We  call the character $P_\lambda$, 
which is independent of $a$,  
{\it multiplicity} of the invariants $N_g(\lambda + aF)=
N_g(\omega(\lambda+a F))$.

$$
\vbox{\offinterlineskip
\hrule
\halign{ \strut 
\vrule#    
& $\;$ \hfil #  \hfil   
&&\vrule#  
& $\;$ 
   # \hfil        
& \hfil # \hfil        
\cr 
& n=1 
&& (0;0,0,0,0,0,0,0,0,-1)=$\lambda_0$  & $g$=0 &\cr 
\noalign{\hrule}
& n=2 
  && (1;1,0,0,0,0,0,0,0,0)=$\lambda_1$  &  $g$=0 &\cr
& && (3;1,1,1,1,1,1,1,0,0)=$\lambda_7$  &  $g$=1 &\cr
& && (6;2,2,2,2,2,2,2,2,0)=$2\lambda_0$  &  $g$=2 &\cr
\noalign{\hrule}
& n=3 
  && (1;0,0,0,0,0,0,0,0,0)=$\lambda_8$  & $g$=0 &\cr
& && (3;1,1,1,1,1,1,0,0,0)=$\lambda_6$  & $g$=1 &\cr
& && (4;2,1,1,1,1,1,1,1,0)=$\lambda_0+\lambda_1$  & $g$=2 &\cr
& && (6;2,2,2,2,2,2,2,1,0)=$\lambda_0+\lambda_7$  & $g$=3 &\cr
& && (9;3,3,3,3,3,3,3,3,0)=$3\lambda_0$  & $g$=4 &\cr 
& && & & \cr
}
\hrule} 
\,
\vbox{\offinterlineskip
\hrule
\halign{ \strut 
\vrule#    
& $\;$ \hfil #  \hfil   
&&\vrule#  
& $\;\;$ 
        # \hfil        
& \hfil # \hfil        
\cr 
& n=4 
  && (2;1,1,0,0,0,0,0,0,0)=$\lambda_2$ & $g$=0 &\cr 
& && (3;1,1,1,1,1,0,0,0,0)=$\lambda_5$ & $g$=1 &\cr 
& && (4;2,1,1,1,1,1,1,0,0)=$\lambda_1+\lambda_7$ &  $g$=2 &\cr 
& && (4;1,1,1,1,1,1,1,1,0)=$\lambda_0+\lambda_8$ &  $g$=3 &\cr 
& && (5;3,1,1,1,1,1,1,1,1)=$2\lambda_1$ &  $g$=3 &\cr 
& && (6;2,2,2,2,2,2,2,0,0)=$2\lambda_7$ & $g$=3 &\cr 
& && (6;2,2,2,2,2,2,1,1,0)=$\lambda_0+\lambda_6$ &  $g$=4 &\cr 
& && (7;3,2,2,2,2,2,2,2,0)=$2\lambda_1$ &  $g$=5 &\cr 
& && (9;3,3,3,3,3,3,3,2,0)=$2\lambda_7$ &  $g$=6 &\cr 
& && (12;4,4,4,4,4,4,4,4,0)=$4\lambda_0$ & $g$=7 &\cr 
}
\hrule}  
$$
{\leftskip1cm\rightskip1cm
{\bf Table 1}. Minimal dominant weights in $\mathcal P_{+,n}^{min}$ 
up to level $n=4$. 
$(d;a_1,a_2,\cdots,a_9)$ represents the minimal domionant weight 
$\lambda=d H -a_1e_1-\cdots -a_9 e_9$. We also list the arithmetic 
genus $g={1\over2}( (\lambda,\lambda)+2-(\lambda,F)) 
= {(d-1)(d-2)\over 2} -\sum_{i=1}^9 {a_i(a_i-1) \over2}$.  \par}

\vskip0.5cm

\subsection{The multiplicity functions $P_\lambda$}

The multiplicity $P_{\lambda}$ determines corresponding multiplicity 
function $P_{\lambda}(\tau,u_1,\cdots,u_8)$ when we evaluate it 
by $u_1 \alpha_1+\cdots+u_8\alpha_9+\tau(e_9+F) \in H^2(S,\bold Z)$, 
i.e.,
$$
P_\lambda(\tau,u_1,\cdots,u_8):= 
P_\lambda( u_1 \alpha_1+\cdots+u_8\alpha_9+\tau(e_9+F) ).
$$
As we observe in (\ref{eqn:ZandP}), 
there is a similarity between $P_\lambda$ and the 
numerater of the Weyl-Kac character formula for the integrable  
representation of affine Kac-Moody algebra[Kac]. As in the case 
for the Weyl-Kac character formula, we may write the multiplicity 
functions, at least formally, in terms of the theta function of 
the $E_8$ lattice. 

\begin{proposition} \thlabel{th:Po} 
For $n \lambda_0 \in \mathcal P_{+,n}^{min}$, we have 
$$
P_{n\lambda_0}(\tau,u_1,\cdots,u_8)=\Theta_{E_8}(n\tau,n u_1, \cdots, n u_8) 
\;\;,
$$
where 
$\Theta_{E_8}(\tau,u_1,\cdots,u_8)=\sum_{l \in E_8} 
e^{2\pi \sqrt{-1} 
\left( \frac{(l,l)}{2}\tau + (l, u_1\alpha_1+\cdots+u_8\alpha_8)
\right)} $ is the theta function of the $E_8$-lattice. 
\end{proposition} 

\begin{proof} The affine Weyl group $W_{\hat E_8}$ is represented by 
a semi-direct product of the translation group $T=\{t_\gamma | \gamma 
\in E_8(-1) \}$ and the 
classical Weyl group generated by $s_{\alpha_1},\cdots,s_{\alpha_8}$. 
Since the classical Weyl group is exactly the stabilizer of $n \lambda_0 \in 
\mathcal P_{+,n}^{min}$, we have from (\ref{eqn:ZandP}),
\begin{align*}
P_{n\lambda_0}&=\sum_{\omega\in W_{\hat E_8}(n\lambda_0)} 
e^{2\pi \sqrt{-1}(\omega(n \lambda_0)-n\lambda_0)} 
=\sum_{\gamma \in E_8(-1)} 
e^{2\pi \sqrt{-1}(t_\gamma(n\lambda_0)-n\lambda_0)}     \\
&= \sum_{\gamma \in E_8(-1)} 
e^{2\pi \sqrt{-1}\left(n\gamma-\frac{(\gamma,\gamma)}{2}n F \right) } \;\;. 
\end{align*}
Evaluating the character with $u_1\alpha_1+\cdots+u_8\alpha_8 
+\tau(e_9+F)$, we obtain the desired result. 
\end{proof}

Explicit form of the function $P_\lambda(\tau,u_1,\cdots,u_2)$ 
for general $\lambda$ contains summation over an non-trivial 
group $W_{\hat E_8}(\lambda)$ and complicated in general. However 
for lower levels $n$ and special values for $u_1,\cdots,u_8$, 
we may have simple form for the multiplicity function. 
For example, in case of $n=2$, we have three elements $2\lambda_0, 
\lambda_1, \lambda_7$ in $\mathcal P_{+,2}^{min}$, and the multiplicity 
functions
$$
P_0(\tau):=P_{2\lambda_0}(\tau(e_9+F))\;\;,\;\;
P_{even}(\tau):=P_{\lambda_1}(\tau(e_9+F))\;\;,\;\;
P_{odd}(\tau):=P_{\lambda_7}(\tau(e_9+F)),
$$ 
have the following simple forms, which were first appeared in [MNVW][Yo]. 

\begin{proposition} \thlabel{th:PePo} ([MNVW],[Yo]) 
For the multiplicity functions defined above, we have; 
\begin{align*}
&P_{even}(\tau)=\left( \frac{E_4(\tau) + E_4(\tau+\frac{1}{2})}{2} 
-E_4(2\tau) \right) q^{-1} \;\;,\;\;  \\
& P_{odd}(\tau)=\left( \frac{E_4(\tau) - E_4(\tau+\frac{1}{2})}{2} \right) 
q^{-\frac{1}{2}} \;\;,\;\; P_0(\tau)=E_4(2\tau) \;\;,\;\; 
\end{align*}
where $E_4(\tau)$ is the Eisenstein series of weight four which 
is a special value of $\Theta_{E_8}$, i.e. $E_4(\tau)=\Theta_{E_8}(
\tau,0,\cdots,0)$. 
\end{proposition}

Since derivation of these forms, and further generalizations to $n=3$,  
from our definition (\ref{eqn:ZandP}) are easy, we do not reproduce them 
here.

\subsection{ Theta function $\Theta_{E_8}$ } Here we summarize a convenient 
realization of the theta function $\Theta_{E_8}(\tau,u_1,\cdots, u_8)$, 
which are often used in the literatures. To  do this, let us consider  
$\bold R^9$ with its orthonormal basis $\varepsilon_1,\cdots,\varepsilon_9$, 
(,$(\varepsilon_i,\varepsilon_j)=\delta_{ij}$). In this space we realize 
the $E_8$ lattice $\sum_{i=1}^9 \bold Z \alpha_i$ by setting $\alpha_1=
\frac{1}{2}(\varepsilon_1-\varepsilon_2-\cdots-\varepsilon_7+\varepsilon_8)$, 
$\alpha_i=\varepsilon_i-\varepsilon_{i-1} \;(2\leq i\leq 7)$, 
$\alpha_8=\varepsilon_1+\varepsilon_2$. Then the $E_8$ theta function 
may be evaluated to 
\begin{equation}
\Theta_{E_8}(\tau,u_1,\cdots,u_8)=\sum_{\gamma \in E_8} 
e^{2\pi \sqrt{-1}\left( \frac{(\gamma,\gamma)}{2} + (\gamma, 
u_1\alpha+\cdots+u_8\alpha_8) \right)} 
=\frac{1}{2}\sum_{i=1}^4 \prod_{j=1}^8 \theta_l(\tau,z_j) \;,
\mylabel{eqn:EeightTheta}
\end{equation}
with 
\begin{align*}
&
\theta_1(\tau,z) :=
i \sum_{n\in \bold Z}(-1)^n q^{(n+{1\over2})^2} y^{n+{1\over2}} \;,\;
&&
\theta_3(\tau,z):=\sum_{n\in\bold Z} q^{n^2} y^n \;,\;  \hfill 
\cr
&
\theta_2(\tau,z):=\sum_{n\in \bold Z} q^{(n+{1\over2})^2} y^{n+{1\over2}} 
\;,\; 
&&
\theta_4(\tau,z):=\sum_{n\in\bold Z}(-1)^n q^{n^2} y^n \;,\;\;\;  
\end{align*}
where $q=e^{2\pi\sqrt{-1}\tau}, y=2\pi \sqrt{-1}z$ and $z_1,\cdots,z_8$ 
are determined by the relation $\sum_{i=1}^8 u_i\alpha_i=\sum_{j=1}^8 
z_j \varepsilon_j$. Hereafter we denote the right hand side of 
(\ref{eqn:EeightTheta}) 
by $\Theta_{E_8}^{\bold Z}(\tau,z_1,\cdots,z_8)$. Namely, 
$\Theta_{E_8}^{\bold Z}(\tau,z_1,\cdots,z_8)$ and 
$\Theta_{E_8}(\tau,u_1,\cdots,u_8)$ should be related by the linear 
relation $\sum_{i=1}^8 u_i\alpha_i=\sum_{j=1}^8 z_j \varepsilon_j$.


\section{Orbit decomposition and BPS numbers}

In this section we study the solutions of the holomorphic anomaly equation 
(\ref{eqn:HAEa}). So far we do not have general proof about that our  
holomorphic anomaly equation (\ref{eqn:HAEa}) really evaluates the generating 
function of the Gromov-Witten invariants defined by (\ref{eqn:GWdef}), 
although we may verify that it produces consistent predictions 
$N_g(\beta)$ for many $\beta$. 
Under this circumstance, our approach is {\it to assume} that 
the generating function defined in (\ref{eqn:characterZ}),  
or more precisely $\mathcal Z_{g;n}(\tau(e_9+F))$, is a solution of  
the holomorphic anomaly equation (\ref{eqn:HAEa}).

\subsection{Vanishing conditions} 
As we see in (\ref{eqn:HAEa}), in order to solve the holomorphic 
anomaly equation we need to fix ``integration constants 
$f_{2g+6n-2}(E_4,E_6)$'', the polynomial ambiguity which 
appears in the integration. 
This polynomial ambiguity is sometimes called {\it holomorphic ambiguity} 
in literatures. We see that the following  requirements for Gromov-Witten 
invariants (BPS numbers) provide conditions to fix this ambiguity. 
The meaning of BPS numbers will be summarized briefly in section 3.2.

\begin{definition} \thlabel{th:vanishing} 
(Vanishing conditions on BPS numbers) 
We define the BPS number $n_h(\beta)$, for $\beta$ satisfying 
$(\beta,F)\geq 1$, by the relation (\ref{eqn:GVformula}). Then we 
impose $n_h(\beta)=0$ unless the following conditions are satisfied:
\item{(i)} $d\geq 1, a_1,\cdots,a_9\geq0$ for $\beta=dH-a_1e_1-\cdots-a_9e_9$ 
if $(\beta,F) \geq 2$, 
\item{(ii)} $\beta=e_i \;(i=1,\cdots,9)$ or 
$d\geq 1, a_1,\cdots,a_9\geq0$ if $(\beta,F)=1$, 
\item{(iii)} $0 \leq h \leq \frac{1}{2}\{ (\beta,\beta)-(\beta,F)+2 \}$. 
\end{definition}

In order to impose the vanishing conditions on $Z_{g;n}$, it is 
useful to introduce the following notations (with $q=e^{2\pi \sqrt{-1}\tau}$):
\begin{equation*}
\tilde Z_{h;n}(q):= \sum_{\beta \in H^2(S,\bold Z), (\beta,F)=n} 
n_h(\beta) q^{(\beta,e_9+F)}\;\;,
\end{equation*}
which is related to $Z_{g;n}(q)$  by 
$$
Z_{g;n}(q)= \sum_{k|n} k^{2g-3} \sum_{h=0}^g C(h,g-h) \tilde Z_{h;n/k}(q^k) 
\;\;.
$$
Since from the defining relation (\ref{eqn:GVformula}), we have $n_h(\beta)
=n_h(w(\beta)) \; (w\in W_{\hat E_8})$ and therefore 
we may consider the orbit decomposition $\tilde Z_{g;n}(q)=\sum_{\lambda \in 
\mathcal P_{+,n}^{min}} \tilde Z_{h;\lambda}(q) P_{\lambda}(q)$ in a similar   
way to $Z_{g;\lambda}(q)$. In this case, the function 
$\tilde Z_{h,\lambda}(q)$ have the following form,
\begin{equation}
\tilde Z_{h,\lambda}(q)=
\sum_{a\geq a_0} n_{h}(\lambda+a\,F) q^{(\lambda+a\,F, e_9+F)} \;\;.
\mylabel{eqn:tildeZlam}
\end{equation}
Here we note that, from the vanishing conditions (i),(ii) and the definition 
of the minimal dominant weights $\lambda \in \mathcal P_{+,n}^{min}$, the 
sum over $a\geq a_0$ is in fact restricted to $a\geq a_0\geq 0$. 
Then since $(\lambda+aF,e_9+F)=(\lambda,e_9)+n+a \geq n$ for 
$\lambda\not=e_9$, we see that  $\tilde Z_{h,\lambda}(q)$ starts 
from an order higher than $q^n$ for $\lambda\not=e_9$. 
( The case $\lambda=e_9 \in \mathcal P_{+,n}^{min}$ is 
possible only for $n=1$. For simplicity, we omit this case from our 
consideration in what follows.) Now since 
$P_{\lambda}(q)=1+\text{(higher order terms in } q)$  by 
(\ref{eqn:ZandP}), we see that 

$(*)$ 
{\it For $n\geq 2$ the $q$-expansion of $\tilde Z_{h,n}(q)$ starts 
from an order higher than $q^n$. } 
 
\noindent 
This is an easy way to impose the vanishing condition (i),(ii), and 
is equivalent to the {\it gap condition} imposed for $Z_{g=0,n}$ in 
[MNVW].  The third condition (iii) further restricts the lower bound 
$a_0$ in (\ref{eqn:tildeZlam}) depending $h$,  and as a result, 
we have much refined conditions for the $q$-expansion of $\tilde Z_{h,n}(q)$. 
Since the arguments are straightforward, we omit its details here.  

The vanishing condition (*) and its refinement with the condition (iii) 
are those what we have in order to fix the ``integration constants'' 
$f_{2g+6n-2}(E_4,E_6)$.  In the case of $g=0$, the conditions from the 
vanishing condition (*) grow linearly in $n$ whereas the dimensions 
of the integration constants $f_{6n-2}(E_4,E_6)$, i.e. dimensions 
of modular forms of weight $6n-2$, do not.  Therefore the existence 
of the solution satisfying the vanishing condition is highly non-trivial. 
In ref.[MNW], the existence was shown by constructing the solutions 
explicitly for $g=0$.  This situation is similar for our higher genus 
generalization (\ref{eqn:HAEa}). However the corresponding 
explicit closed formula of the solutions has been obtained only for $g=1$. 
For $g\geq2$, the existence 
of the solution satisfying the vanishing conditions are verified for 
lower values of $g$ and $n$, e.g. $g, n\leq 10$.  Some of them are 
displayed in the end of the section 1.

\subsection{Orbit decomposition $\mathbf{n\leq2}$}

The case for $n=1$, the orbit decomposition is rather trivial 
since the set $\mathcal P_{+,1}^{min}$ consists only one element 
$\lambda_0$. Then, for example, the initial data $Z_{0,1}(\tau)=
\frac{q^{\frac{1}{2}} E_4(q)}{\eta(q)^{12}}$ in (\ref{eqn:HAEa}) 
is decomposed to 
$$
Z_{0,1}(\tau) = \frac{q^{\frac{1}{2}}}{\eta(q)^{12}} 
P_{\lambda_0}(\tau,0,\cdots,0) 
\;\;,
$$
where, by Proposition \ref{th:Po}, $P_{\lambda_0}(\tau,0,\cdots,0)=
\Theta_{E_8}(\tau,0,\cdots,0)=E_4(q)$. This implies that 
$$
\mathcal Z_{0,\lambda_0}(\tau(e_9+F))=\sum_{a\geq0} N_0(\lambda_0+a\,F) 
q^{(\lambda_0+a\,F,(e_9+F))}
={1 \over \prod_{m >0}(1-q^m)^{12}} \;\;,
$$
which is in the same form, except the power 12 replaced by 24, 
as the counting function for the nodal rational curves in K3 surfaces 
found in [YZ]. See [BL] and also [HSS],[HST1,2] for detailed interpretations. 
For higher genus, $Z_{g;1}(\tau)$, the orbit decompositions are 
simply achieved dividing by the multiplicity function $P_{\lambda_0}(\tau)
=E_4(\tau)$, i.e., we simply have $Z_{g,\lambda_0}(\tau)=Z_{g;1}(\tau) 
(P_{\lambda_0}(\tau))^{-1}$. 

For the level $n=2$ cases, we need to make the following decomposition,
\begin{equation*}
Z_{g;n}(\tau)=Z_{g,2\lambda_0}(\tau)P_{2\lambda_0}(\tau) 
+ Z_{g,\lambda_{even}}(\tau)P_{\lambda_{even}}(\tau) 
+ Z_{g,\lambda_{odd}}(\tau)P_{\lambda_{odd}}(\tau) \;,
\end{equation*}
where $\lambda_{even}=\lambda_1, \lambda_{odd}=\lambda_7$ (, see 
Table 1). This decomposition has been done for $g=0$ in [MNVW][Yo] noticing  
modular properties of the functions $P_{2\lambda_0}(\tau), 
P_{\lambda_{even}}(\tau)$ and $P_{\lambda_{odd}}(\tau)$, e.g. 
$P_{2\lambda_0}(\tau)=E_4(2\tau)$ is a modular form of the group 
$\Gamma_1(2)$. Since $E_2$ does not behave modular form, the $E_2$-dependence 
of $Z_{g;n}(\tau)$  should be found in $Z_{g,\lambda}(\tau)$. 
Then using the identity 
\begin{align*}
P_{\lambda_0}(\tau,u_i)^2 = 
P_{\lambda_0}(\tau,0)P_{2\lambda_0}(\tau,u_i) 
 + C_{\lambda_{even}} 
P_{\lambda_{even}}(\tau,0)
&P_{\lambda_{even}}(\tau,u_i)  \\ 
 +  C_{\lambda_{odd}}
P_{\lambda_{odd}}(\tau,0)
&P_{\lambda_{odd}}(\tau,u_i) , 
\end{align*}
and linear independence of $P_{\lambda}(\tau)$'s, 
we may derive the holomorphic anomaly equation for 
$Z_{g,\lambda}(\lambda \in \mathcal P_{+,n}^{min})$;
\begin{equation}
\frac{\pd Z_{g,\lambda}(\tau)}{\pd E_2} 
={C_{\lambda} \over 24}\sum_{g'+g''=g} 
Z_{g',\lambda_0}(\tau)
Z_{g'',\lambda_0}(\tau)  
P_{\lambda}(\tau,0)
+{1\over 4} Z_{g-1,\lambda} \;\;,
\mylabel{eqn:HAEorbit}
\end{equation}
where $C_{\lambda}=1,\frac{q^2}{135}, \frac{q}{120}$, respectively, 
for $\lambda=2\lambda_0,\lambda_{even}, \lambda_{odd}$. Integrating 
(\ref{eqn:HAEorbit}) for $g=0$, in [MNVW] and [Yo] the following 
forms are determined;
\begin{align*}
&
Z_{0,2\lambda_0}(\tau)={1\over24}{q \over \eta(\tau)^{24}} 
\big\{ {1\over16}(4 G_2(\tau)^2-3G_4(\tau))E_2(\tau)+
      {1\over8}(2 G_2(\tau)^2-3G_4(\tau)) G_2(\tau) \big\} \cr 
&
Z_{0,\lambda_{even}}(\tau)={1\over24}{q^2 \over \eta(\tau)^{24}} 
G_4(\tau)
\big\{ {1\over16}E_2(\tau)-
      {1\over8}G_2(\tau) \big\} \cr 
&
Z_{0,\lambda_{odd}}(\tau)={1\over24}{q^{{3\over2}}\over \eta(\tau)^{24}} 
G_4(\tau)^{1\over2}
\big\{ {1\over16}G_2(\tau)E_2(\tau)-
      {1\over32}(2G_2(\tau)^2+3G_4(\tau)) \big\}\;, 
\end{align*}
where $G_2(\tau):=\theta_3(\tau,0)^4+\theta_4(\tau,0)^4, \; 
G_4(\tau):=\theta_2(\tau,0)^8$ are generators of the ring of 
the modular forms of $\Gamma_1(2)$. Now their argument extends straight 
forward way to our cases $g\geq 2$. The results are as follows:

\begin{proposition} 
The characters $Z_{g,\lambda}(\tau (F+e_9)) \;  
(\tilde \lambda \in P^{min}_{+,2})$ may be written in terms of 
the generators $G_2(\tau), G_4(\tau)$ of the modular forms of $\Gamma_1(2)$. 
\end{proposition}

Here we list the results up to $g=3$, although calculations continues to 
higher $g$ as well:  

\vskip0.2cm

\noindent
\underline{(i) $2\lambda_0$}
$$
\begin{aligned}
Z_{1,2\lambda_{0}}(\tau)=&
{1\over 24^2 64}{q\over \eta^{24}} 
(20 (4G_2^2-3G_4)E_2^2 +48(2G_2^3-3G_2G_4)E_2      & \cr
& \hskip4cm  +28G_2^4 -27 G_2^2 G_4 + 27 G_4^2  )  & \cr
Z_{2,2\lambda_{0}}(\tau)=&
{1\over 24^4 20}{q \over \eta^{24}} 
\big( 380 (4G_2^2-3G_4)E_2^3
+1080(2G_2^3-3G_2G_4)E_2^2                        & \cr 
& +3(428G_2^4-387 G_2^2 G_4 + 387 G_4^2)E_2 
+356 G_2^5 - 636 G_2^3 G_4 - 432 G_2 G_4^2 \big)  & \cr
Z_{3,2\lambda_{0}}(\tau)=&
{1\over 24^5 1120}{q\over \eta^{24}} 
\big( 36400 (4G_2^2-3G_4)E_2^4
+120960(2G_2^3-3G_2G_4)E_2^3                   & \cr
&+4200(52G_2^4-45 G_2^2 G_4 + 45 G_4^2)E_2^2   & \cr
&+64(1873G_2^5-3378 G_2^3 G_4 -2241 G_2 G_4^2)E_2  & \cr
&+30444 G_2^6 - 54117 G_2^4 G_4 + 113454 G_2^2 G_4^2 +31995 G_4^3 \big)  &\cr
\end{aligned}
$$
\noindent
\underline{(ii) $\lambda_{even}=\lambda_1$ }  
$$
\begin{aligned}
Z_{1,\lambda_{even}}(\tau)&=
{1\over 24^2 64}{q^2 \over \eta^{24}} 
G_4\left( 20 E_2^2-48 G_2 E_2 + 13G_2^2+15 G_4\right) 
\cr
Z_{2,\lambda_{even}}(\tau)&=
{1\over 24^4 20}{q^2 \over \eta^{24}} G_4\big( 
380 E_2^3 -1080 G_2 E_2^2 + 3 (197G_2^2+231 G_4) E_2   \cr 
&\hskip6cm   
- 4(25G_2^3+153 G_2G_4) \big) 
\cr
Z_{3,\lambda_{even}}(\tau)&=
{1\over 24^5 1120}{q^2 \over \eta^{24}} G_4\big( 
36400 E_2^4 -120960 G_2 E_2^3 + 840 (119G_2^2+141 G_4) E_2^2  \cr
& 
- 128(262 G_2^3+1611 G_2G_4)E_2 
+ 3(1301 G_2^4+27726 G_2^2 G_4 +11565 G_4^2) \big) 
\end{aligned}
$$

\noindent
\underline{ (iii) $\lambda_{odd}=\lambda_7$ }

\begin{align*}
Z_{1,\lambda_{odd}}(\tau)&=
{1\over 24^2 64}{q^{3\over2}\over \eta^{24}} 
G_4^{{1\over2}}\left( 20 G_2 E_2^2-12 (G_2^2+3G_4) E_2 +G_2^3 +27 G_2G_4 
\right) 
\cr
Z_{2,\lambda_{odd}}(\tau)&=
{1\over 24^4 40}{q^{3\over2}\over \eta^{24}} 
G_4^{{1\over2}}\big( 760 G_2 E_2^3
-540(G_2^2+3G_4) E_2^2 
+6(17G_2^3+411G_2G_4) E_2  \cr 
&\hskip6cm 
-11G_2^4 -846 G_2^2G_4 -567 G_4^2 \big) 
\cr
Z_{3,\lambda_{odd}}(\tau)&=
{1\over 24^5 1120}{q^{3\over2}\over \eta^{24}} 
G_4^{{1\over2}}\big( 36400 G_2 E_2^4
-30240(G_2^2+3G_4) E_2^3    \\
&
+840(11G_2^3+249G_2G_4) E_2^2   
-8(223G_2^4+17838 G_2^2G_4+11907 G_4^2)E_2   \\
& +3(29G_2^5+10206 G_2^3G_4+30357 G_2G_4^2) \big) 
\cr
\end{align*}

{\bf Remark.} Since the weights $\lambda_{even}$ and $\lambda_{odd}$ 
are primitive, we have 
$$
Z_{g,\lambda_{even}}(q)=\sum_{h=0}^g C(h,g-h) \tilde Z_{h,\lambda_{even}}(q) 
\;\;,
$$
and the corresponding formula for $\lambda_{odd}$.

\subsection{BPS numbers $n_g(\beta)$} 
The BPS numbers $n_h(\beta)$ 
are related to Gromov-Witten invariants $N_g(\beta)$ by the formula 
(\ref{eqn:GVformula}). When $g=0$ this formula reduces to 
$N_0(\beta)=\sum_{k|\beta} \frac{1}{k^3} n_0(\beta/k)$, 
which appeared in the original work by  Candelas, de la Ossa, Green 
and Park [CdOGP] where it was found that $n_0(\beta)$ is integer-valued 
and interpreted  as the number of rational curves of a fixed homology 
class $\beta$. When the rational curves are smooth and isolated, i.e. 
$O_C(-1)\oplus O_C(-1)$ curves in Calabi-Yau threefolds $X$, 
it is natural to have $n_0(\beta)=1$, and in this case 
the multiple cover formula was proved in [AM][Ma]. 
Also, in this case, the higher genus generalization 
(\ref{eqn:GVformula}) was proved [FP] under further assumption that 
$\beta$ is primitive. (In [FP], the formula (\ref{eqn:GVformula}) 
was proved also for the case $\beta$ represents a super-rigid 
elliptic curve.)  

Gopakumar-Vafa conjecture mentioned in section 1 contains 
a proposal for a ``definition'' of the number 
$n_h(\beta)$, which is independent to Gromov-Witten theory.  
The idea from string theory is that we may regard the number $n_g(\beta)$  
as the number of BPS states of spin $g$ and charge $\beta$ in 
the context of M-theory. 
To describe its mathematical aspects briefly following [HST2],  
let $X$ be a Calabi-Yau threefolds with an ample divisor $L$, and 
consider a moduli space $\mathcal M_\beta(X)$ of D2-branes, certain 
local systems supported on curves with homology class $\beta$. 
Under a suitable stability condition via $L$, the moduli space 
$\mathcal M_\beta(X)$ becomes projective. In [HST2], it is found 
that fixing the Hilbert polynomial to $P(m)=d m+1$ $(d=\beta\cdot L)$ 
gives rise a moduli space consistent to the expectation from physics. 
We may consider a natural map $\pi_\beta: \mathcal M_\beta(X) 
\rightarrow Chow_\beta(X)$, where $Chow_\beta(X)$ is a subvariety 
in the Chow variety $Chow_d(X)$ of degree $d$. Writing 
$S_\beta=\pi_\beta(\mathcal M_\beta(X))$ 
we have a surjective morphism $\pi_\beta: \mathcal M(X) \rightarrow S_\beta$. 
This is a brief sketch of the mathematical definitions made in [HST2] 
for the moduli spaces of D2 branes. 
Gopakumar and Vafa futher {\it expect} that 
there exist two Lefshetz $sl_2$'s which act  
on the cohomology space $H^*(\mathcal M_\beta(X))$,  one from 
the fiberwise Lefshetz action, denoted by $sl_{2,L}$, and the other from 
that of the base $S_\beta$, and denoted by $sl_{2,R}$. 
They also expects these two $sl_2$ commute and act on the $E_2$-term of 
the Leray spectral sequence. 
In [HST2], it has been pointed out that to ensure these $sl_2$ actions 
of the desired properties we need to use the Leray spectral sequence 
of the perverse sheaves[BBD] to the morphism $\mathcal M_\beta(X) 
\rightarrow S_\beta$. In this case, the sequence degenerates at the 
$E_2$-term and the two commuting $sl_2$ actions are realized in the 
interesection cohomology ring of $\mathcal M_{\beta}(X)$.

Assuming their existence, although the exsistence should be ensured as above,  
Gopakumar and Vafa has identified these 
two Lefschetz $sl_2$ actions on 
$H^*(\mathcal M_\beta(X))$ with the 
spin operators $SU(2)_L\times SU(2)_R$ acting on the BPS states in 
5 dimensions. Then they introduce the following decomposition 
(in the representation ring);
\begin{equation}
H^*(\mathcal M_{\beta}(X))=
(I_0\otimes R_0) \; \oplus \;  
(I_1\otimes R_1) \; \oplus \;
\cdots \; \oplus \; (I_g\otimes R_g)
\mylabel{eqn:GVdecomp}
\end{equation}
where $I_h=\left( (0)\oplus(\frac{1}{2}) \right)^{\otimes h}$ is the 
$sl_{2,L}$ representation.  
The $sl_{2,R}$ representation $R_h (0\leq h \leq g=g_\beta)$ should 
be understood as defined by the above decomposition. 
Then the invariants $n_h(\beta)$, which are integral by definition, 
are given by the ``index'';
\begin{equation}
n_h(\beta)=Tr_{R_h} (-1)^{H_R}  \;\;,
\mylabel{eqn:nBPS}
\end{equation}  
where $H_R$ is the generator of the Cartan subalgebra of $sl_{2,R}$ 
in Chevalley basis. This is the proposed ``definition'' of the number 
of BPS states of spin $h$ and charge $\beta$. This proposed ``definition'' 
has been made mathematically more precise based on the definition 
$\mathcal M_\beta(X)$ and the Leray spectral sequence for perverse 
sheaves as described above. Based on this precise definition, 
the cases in which $\beta$ represents a multiple of a rigid rational curve 
and also a multiple of a super-rigid elliptic curve $E$ are studied 
in details, and consistent answers are obtained for $n_h(\beta)$. 
(See also [BP].) Also a closed formula [HST1, Proposition 1.2, 1.3] for 
$\sum_g Z_{g;1}(q) \lambda^{2g-2}$ has been proved by this precise 
definition [HST2, Theorem 4.10].

If we think that Gopakumar-Vafa conjecture, the formula (\ref{eqn:GVformula}), 
connects the BPS numbers defined above to Gromov-Witten invariants,  
the content of the conjecture becomes highly non-trivial as explained above. 
However, here in this paper, we simply list the results for 
$n_h(\beta)$ which results from Gromov-Witten invariants assuming 
the relation (\ref{eqn:GVformula}). In Table 2 -- 5, we have listed 
the numbers for $n_h(\beta)$ for $n=1,2,3,4$ and for each 
$W_{\hat E_8}$-orbits, (see section 3.1).  As we see in our listing, 
the resulting BPS numbers are all integers supporting Gopakumar-Vafa 
conjecture. Furthermore we may interpret some of these numbers 
following the expected `definition' (\ref{eqn:nBPS}) (, see Remark below). 

To make our listing, we have to accomplish the orbit decompositions 
for higher levels (, $n=3,4$ ). Since the process is so technical, we 
omit the details here. But the idea is to use holomorphic anomaly 
equation (\ref{eqn:HAEa}) for other parametrizations $\mathcal Z_{g;n}(t D)$ 
with $D=H, e_9+F, e_8+e_9+F, e_7+e_8+e_9+F, e_6+e_7+e_8+e_9+F$, respectively, 
and make orbit decompositions for each.  For example, the multiplicity 
function $P_{\lambda_0}(tD)$ may be determined to be 
$$
\Theta_{E_8}^{\bold Z}(t;0^8) \;\;,\;\;
\Theta_{E_8}^{\bold Z}(2t;t,t,0^6) \;\;,\;\;
\Theta_{E_8}^{\bold Z}(3t;2t,t,t,0^5) \;\;,\;\;
\Theta_{E_8}^{\bold Z}(4t;3t,t,t,t,0^4) \;\;,
$$
respectively for $D=e_9+F, e_8+e_9+F, e_7+e_8+e_9+F, e_6+e_7+e_8+e_9+F$. 
The form $P_{\lambda_0}(tH)=\Theta_{E_8}^{\bold Z}(3t;t,\cdots,t,-t)$ was 
first appeared in [HSS]. These parametrizations have also been utilized in 
a recent work [Moh]. 

{\bf Remark.} (1) As we observe in our Tables 2--5, the numbers 
$n_h(\beta)$ are {\it integral}. Similar observations are also 
made in [KZ][KKV] for several del Pezzo surfaces in Calabi-Yau 
threefolds.  Since the Gromov-Witten invariants $N_g(\beta)$ are 
invariant under bi-rational transformations (if $\beta$ does not 
intersect with the divisor of the bi-rational maps)[AGM], our $N_g(\beta)$ 
or $n_g(\beta)$ for rational elliptic surface $S$ contain the 
corresponding invariants for all del Pezzo surfaces obtained 
by blowing up $k (\leq 9)$ points.  For example, for the class 
$\beta=H=\lambda_8$ in Table 4 we see the genus zero invariants 
for (local) $\bold P^2$, i.e. $n_0(H)I_0 = 3 I_0$. Also in 
Table 4, we see $n_0(\lambda_6)I_0+n_1(\lambda_6)I_1=27 I_0 -4 I_1$, 
i.e. the invariants for the del Pezzo surface $Bl_6$, see e.g. [KZ]. 
(Note that $\lambda_6=(3;1,1,1,1,1,1,0,0,0)$ may be read as the class 
of the anti-canonical bundle on the cubic surface.)
In a similar way, we may continue our identification or interpretation 
of the numbers $n_h(\beta)$, although complete understanding of these 
numbers is beyond our scope of present paper.  
For the case of 
(local) $\bold P^2$, several numbers $n_h(d):=n_h(d H)$ has been 
verified in [KKV] under suitable `understanding' of Gopakumar-Vafa 
conjecture (, see below). 

(2) In ref.[GV], assuming the fibration $\mathcal M_\beta(X) \rightarrow 
S_\beta$ and the decomposition (\ref{eqn:GVdecomp}), 
it is argued in general that 
\begin{equation}
n_0(\beta)=(-1)^{\text{dim}\mathcal M_\beta(X)}
\chi(\mathcal M_\beta(X)) \;\;,\;\;
n_g(\beta)=(-1)^{\text{dim}S_\beta} \chi(S_\beta) \;\;,
\mylabel{eqn:ngEuler}
\end{equation}
where $\chi$ represents the Euler number. (These equations hold also in the 
formulation via intersection cohomology.)   Also the $D2$ brane moduli 
space $\mathcal M_\beta(X)$ is naively claimed as the Jacobian 
fibration made over the moduli space of curves $C \subset X$ with $[C]=\beta$, 
which we write $S_\beta$. This description of the moduli space 
$\mathcal M_\beta(X)$ is to naive since there appears the cases of 
singular curves or even worse non-reduced curves in the family of the 
curves. However this naive definition provides `nice' (although 
not quite correct in general) intuitions for the 
numbers $n_h(\beta)$. For example, the intuition about $n_0(\beta)$ 
is the Euler number of the locus for the nodal rational curves 
appears on $S_\beta$, which has been justified in [YZ][Be] for $X=$K3. 
This intuition is also naively expected in [GV] for general 
$n_h(\beta)$ $(0\leq h\leq g)$, 
i.e. the numbers are the Euler numbers of the locus on $S_\beta$ 
where nodal (genus $h$) curves appear. Again, because of the possible 
complicated degenerations of the curves, it is known that for this  
intuition to work, we need to take into account some corrections, by hand, 
depending on degeneration type[KKV].  

In our case of curves in a surface $S$, the moduli space $S_\beta$ 
of the curves may be understood as the linear system of the divisor 
class $\beta$ identifying $H_2(S,\bold Z)$ with $H^2(S,\bold Z)$. 
Then the predicted numbers $n_g(\beta)$ in (\ref{eqn:ngEuler}) is, 
up to sign, simply (the dimension of the linear system)+1, which we 
can verify all in our listing. In contrary to this, for the verification of 
$n_0(\beta)=(-1)^{\text{dim }\mathcal M_\beta(X)}\chi(\mathcal M_\beta(X))$, 
we need more a precise definition of the moduli space. 
However if we restrict our attention to 
$\beta$'s which give homology classes of elliptic curve in $S$, we may 
explain the numbers $n_0(\beta)$ from a naive 
definition of $\mathcal M_\beta(X)$ as the Jacobian fibration over the 
linear system $S_\beta$. The homology classes which admit this simple 
interpretation are;  
$$
\beta=3H, 3H-e_1, 3H-e_1-e_2, \cdots, 3H-e_1-e_2-\cdots -e_8 \;\;, 
$$
for all of these we have the arithmetic genus 1. In fact, these 
classes may be regarded as the anti-canonical classes of respective 
del Pezzo surfaces 
$Bl_k$ ($k$ points blow up of $\bold P^2$) and therefore general elements 
of the linear system define an elliptic curve. We can find this kind of 
homology classes in our listing;
\begin{equation}
\begin{aligned}
\lambda_0+F&=(3;1,1,1,1,1,1,1,1,0) 
&\lambda_7&=(3;1,1,1,1,1,1,1,0,0)&    \cr
\lambda_6&=(3;1,1,1,1,1,1,0,0,0) 
&\lambda_5&=(3;1,1,1,1,1,0,0,0,0), &  \cr
\end{aligned}
\end{equation}
and the corresponding numbers $n_0(\beta)I_0 + n_1(\beta)I_1$ are read, 
respectively, as 
\begin{equation}
12 I_0 -2 I_1 \; , \;  -20 I_0 + 3 I_1 \; , \;  
27I_0-4I_1 \;, \; -32 I_0+5 I_1 .
\mylabel{eqn:Is}
\end{equation}
The case $\beta=3H$ is not contained in our listing, since it appears 
in $(\beta,F)=9$, however, it is known the numbers are $27 I_0-10 I_1$ 
(see e.g. [KZ]). In all cases, the number $n_1(\beta)$ is given, 
up to sign, by the dimension of the linear system (plus one) considered 
in the respective del Pezzo surfaces, $Bl_k \; (k=8,7,6,5,0)$. Also we may 
understand the numbers $n_0(\beta)$ following the argument given in [GV] 
for the case $\beta=3H$. Namely, the naive moduli space $\mathcal M_\beta$ 
as the Jacobian fibration may be described by specifying a point on  
curves parametrized by the linear system. Since the specified point can 
move over the respective surface $Bl_k$ $(k\leq 7)$, this entails fibration 
$Bl_k \rightarrow \mathcal M_\beta \rightarrow 
\bold P^{\text{dim }|\beta|-1}$. 
For $k=8$ we need some special cares since the dimension of the 
linear system is one. However, for this case, from slightly different 
view point one may argue that $\mathcal M_\beta = \frac{1}{2}$K3 (, see 
e.g. [HST1,2]). 
Evaluating the Euler number of 
$\mathcal M_\beta$, we obtain $n_0(\beta)=(-1)^{\text{dim} \mathcal M_\beta} 
\chi(Bl_k) \text{dim }|\beta|$ $(k\leq7)$. In this way we explain  
the numbers $n_0(\beta)$ in (\ref{eqn:Is})  as 
$12=\chi(\frac{1}{2}\text{K3}), \;
-20=-\chi(Bl_7)\times \chi(\bold P^1), \;
27=\chi(Bl_6)\times \chi(\bold P^2), \;
-32=-\chi(Bl_5)\times \chi(\bold P^3)$.

Some detailed arguments may be found in [KKV] to `explain' 
the numbers $n_h(\beta)$ as the Euler numbers with some corrections 
of the degeneration locus of curves on $S_\beta$.  
Following the arguments there, we may understand some other 
numbers $n_h(\beta)$ in our tables. 
Recently it is announced to the author that 
for several $\beta$ in the Table 3--5, we can verify $n_h(\beta)$ following 
the definition given in [HST2], i.e. from the definition of 
$\mathcal M_\beta(X)$ given there, the Leray spectral sequence of the 
perverse sheaves and the intersection cohomology[Ta](, Table 2 has 
been proved in [HST2, Theorem 4.10]). However we still do not have 
full geometrical verifications of these integer numbers 
$n_h(\beta)$ presented in Table 3--5.

\vfill\eject

\centerline{\epsfxsize 7.5truein \epsfbox{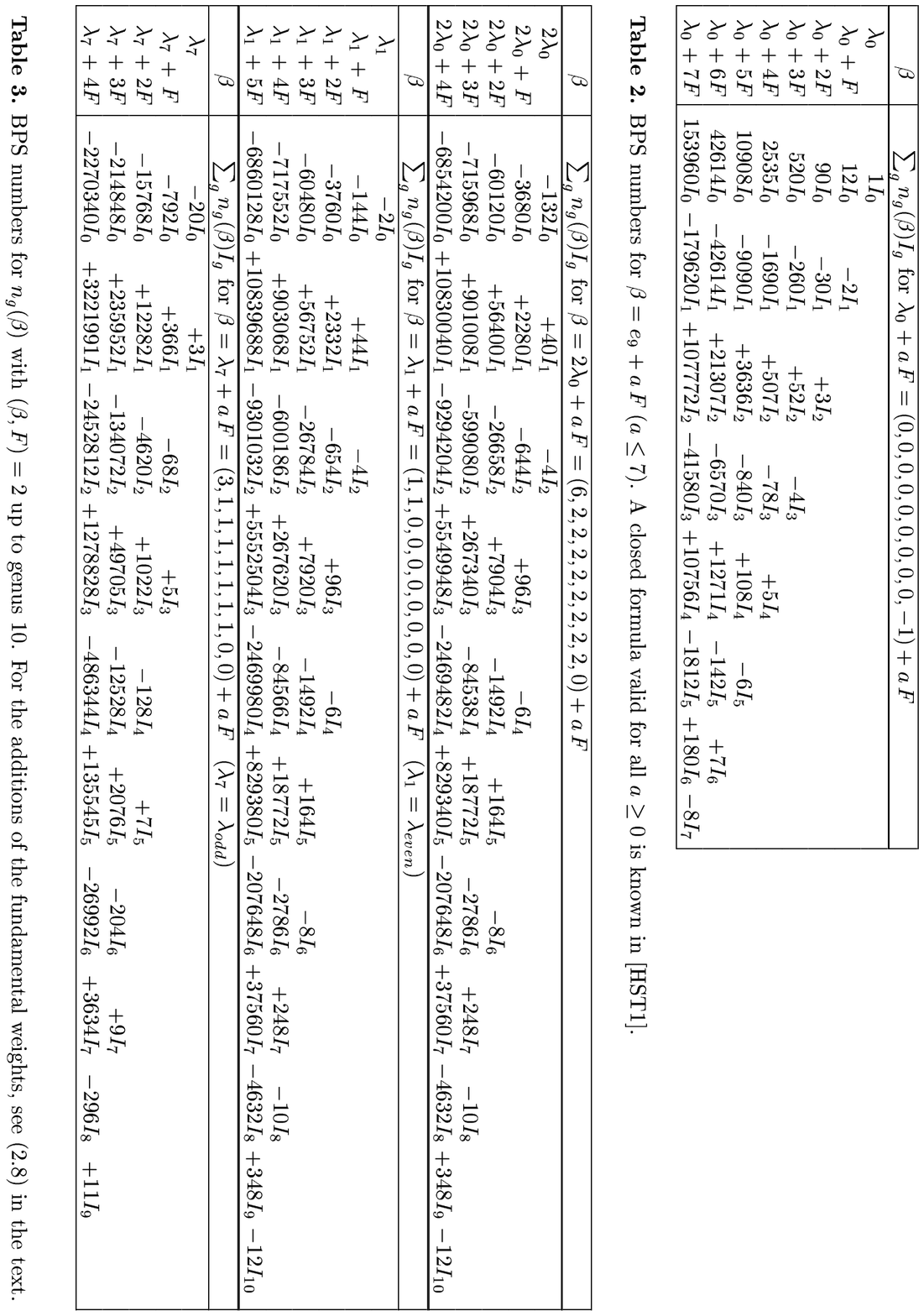}}

\vfill\eject

\centerline{\epsfxsize 7.5truein  \epsfbox{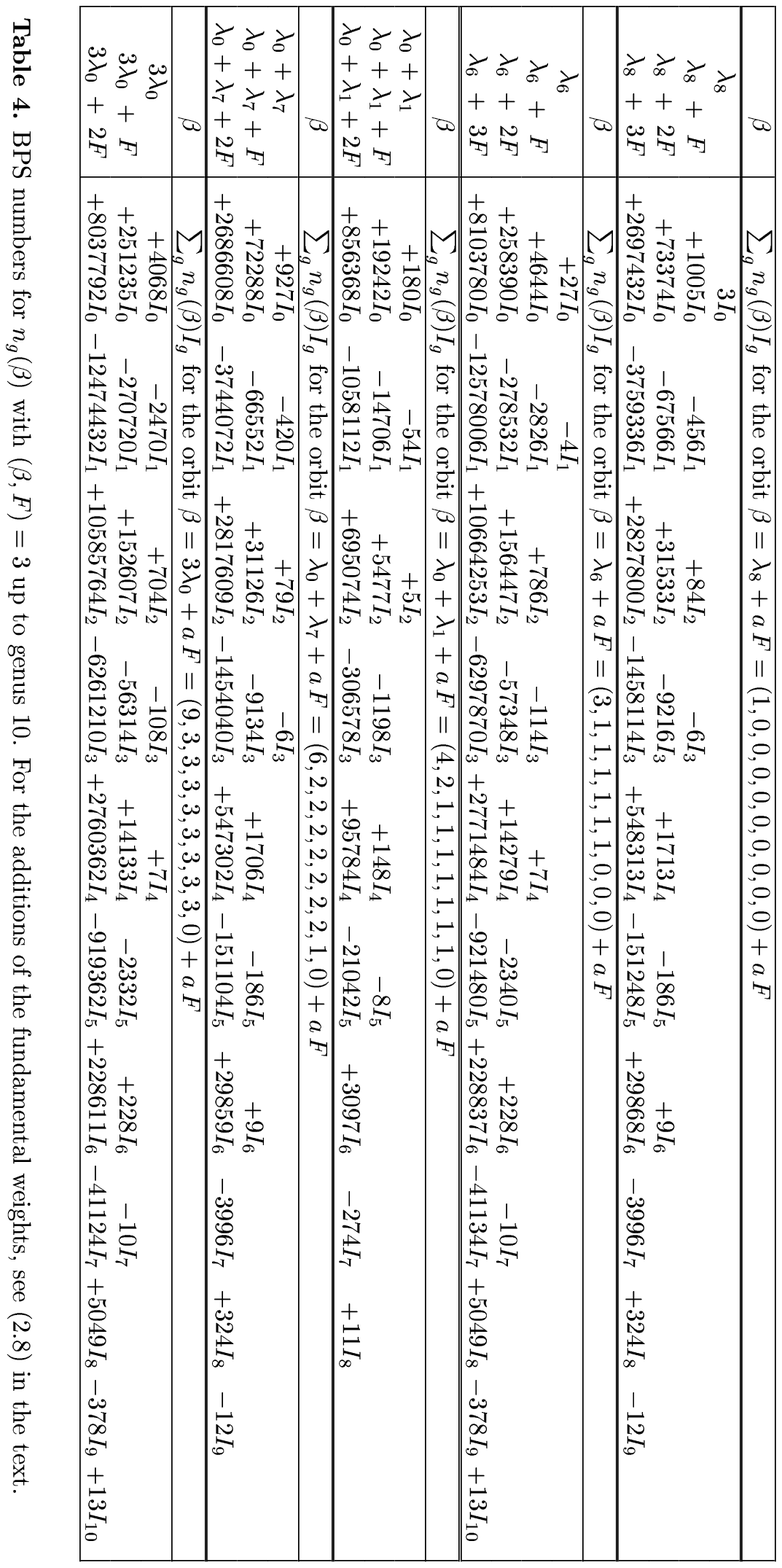}}

\vfill\eject

\centerline{\epsfxsize 7.5truein  \epsfbox{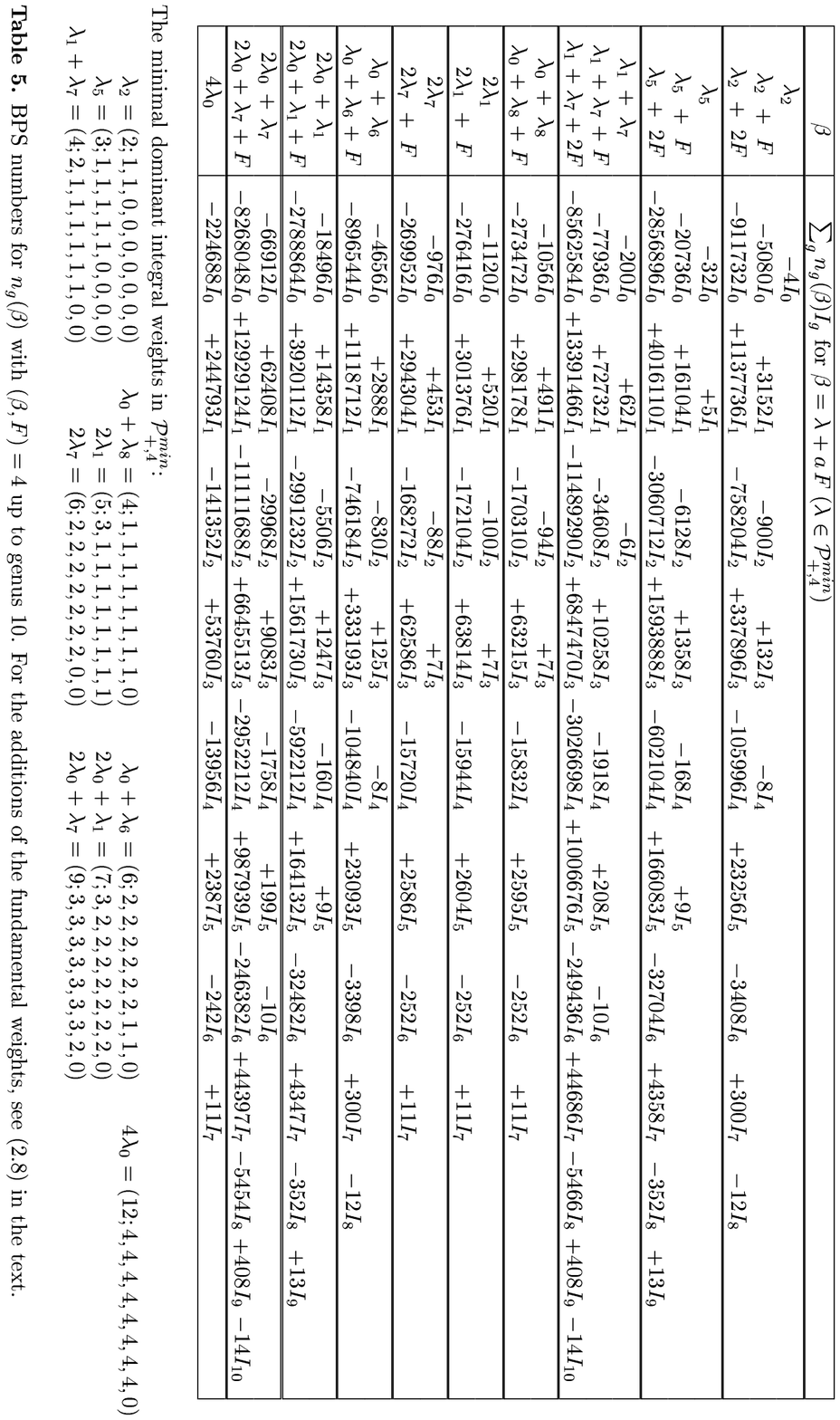}}

\vfill\eject

\section{Bershadsky-Cecotti-Ooguri-Vafa holomorphic anomaly equation}

In previous sections, we have analyzed the holomorphic anomaly equation 
(\ref{eqn:HAEa}) of rational elliptic surface in detail. 
Here we continue our analysis based on Bershadsky-Ceccoti-Ooguri-Vafa 
(BCOV) holomorphic anomaly equation.  BCOV holomorphic anomaly 
equation is a general formula for partition functions 
of the topological sigma model with target space Calabi-Yau 3-folds. 
Therefore it is applicable, in principle, for general Calabi-Yau 3-folds 
to determine the higher genus prepotential $\mathtt F_g$. However, 
unfortunately, solving the equation is so complicated that  Calabi-Yau 
models for which we can determine $\mathtt F_g$ are very restricted (,
e.g. in references [BCOV2][KKV] $\mathtt F_g$ up to  
$g=5$ has been analyzed only for those models of one dimensional 
moduli of K\"ahler deformation, i.e $rk H^2(X,\bold Z)=1$. ). 
In a recent paper [KZ] it has been found that a considerable 
simplification occurs in the local mirror limit finding that the dilaton 
does not propagate under this limit. 
Using this fact prepotentials $\mathtt F_g$ ($g\leq8$) have been determined 
for rational surfaces, $p$-points blow up of $\bold P^2$ ($0\leq p\leq8$) and 
$\bold P^1 \times \bold P^1$, restricting the deformation parameter 
to a specific direction. Although we see considerable simplification 
in the local mirror limit, the higher genus calculations are still 
tedious because of formidable growth of graphs we need to sum up. 

In this section we will analyze the local limit of BCOV holomorphic 
anomaly equation for ${1\over 2}$K3, realizing the surface as a smooth divisor 
in a Calabi-Yau threefold. The aims of this section are two-folded; 
the first is to see a consistency between our equation (\ref{eqn:HAEa}) 
and BCOV holomorphic anomaly equation.  As we will see in the following,  
they produce the same results although their equivalence seems nontrivial.    
The second is to show examples of two parameter deformations 
for which we can still manipulate BCOV holomorphic anomaly equation. 

Recently many progresses have been made in counting holomorphic 
discs, so-called disc instantons, with their boundary on a 
Lagrangian submanifold in (non-compact) Calabi-Yau threefolds. 
See references [OV],[AV],[AVK],[LM], and also [GZ],[LK], 
[LLY2] for suitable extension of the moduli space of stable maps to 
disc instantons.  Most recently, very non-trivial relations to 
Chern-Simons gauge theory which enables us to write down all genus 
generating function has been found in [AMV],[DFG](, e.g. Table 6 in 
[AMV] exactly coincides with our Table 8 below). 
In this paper, however, our attention will be 
restricted to the case of {\it old} instantons.

\subsection{ BCOV holomorphic anomaly equation }

In the original paper by Bershadsky, Cecotti, Ooguri and Vafa [BCOV1], 
the higher genus prepotential $\mathcal F_g$ has been defined as a partition 
function of the topological sigma model with its target space Calabi-Yau 
3 fold $X$ and the world sheet being genus $g$ Riemann surfaces. 
$\mathcal F_g$ is expected to be a holomorphic function (section) on 
the moduli space of Calabi-Yau manifolds after the topological twist, 
however, they found that there is {\it holomorphic anomaly}. To describe 
it very briefly, let us consider a Calabi-Yau threefold $X$, and 
denote its mirror Calabi-Yau threefold by $X^\vee$. 
We consider its (local) deformation family 
$\{ X^\vee_x \}_{x \in \mathcal M^0(X^\vee)}$ writing the deformation 
space by $\mathcal M^0(X^\vee)$. (We are mainly interested in a local 
deformations near so called {\it large complex structure limit}, 
where the monodromy become maximally degenerated.) Since 
the deformations are unobstracted [Ti],[To],[Bo], we may assume 
$\mathcal M^0(X^\vee)$ is smooth, and introduce Weil-Peterson metric 
by the K\"ahler potential $K(x,\bar x)$ with $e^{-K}=\int_{X^\vee} 
\bar\Omega_x \wedge \Omega_x$ where $\Omega_x$ is the nowhere vanishing 
holomorphic 3-from of $X^\vee_x \; (x\in \mathcal M^0(X^\vee))$. 
We may assume a compactified complex structure moduli space 
$\mathcal M^{cpl}(X^\vee)$ in some sense, which naturally exists, e.g. 
for monomial deformations of hypersurfaces, 
and may consider the K\"ahler geometry patching the above local geometry 
on $\mathcal M^{cpl}(X^\vee)$.

Let us denote by $\mathcal L$ the holomorphic line bundle on $\mathcal M^{cpl}
(X^\vee)$ whose section is given by $\Omega_x$.  Then  
$e^{-K(x,\bar x)}$ is a section of $\mathcal L \otimes \bar{\mathcal L}$. 
Also we may consider the Griffith-Yukawa coupling $C_{ijk}:=-\int_{X^\vee} 
\Omega_x \wedge \pd_{x_i}\pd_{x_j}\pd_{x_k} \Omega_x$ and its complex 
conjugate $C_{\bar i \bar j \bar k}:=\overline{ C_{ijk} }$, 
which are regarded as 
a section of $\mathcal L^{\otimes 3}$ and $\bar{\mathcal L}^{\otimes 3}$, 
respectively. BCOV identifies the higher genus prepotential 
$\mathcal F_g$ as an almost holomorphic section of 
$\mathcal L^{2-2g}$ but with holomorphic anomaly described by 
\begin{equation}
\pd_{\bar x_i}\mathcal F_g 
=\frac{1}{2} e^{2K} \sum_{j, k, \bar j, \bar k} 
C_{\bar i \bar j \bar k} G^{j\bar j} G^{k \bar k} 
( \sum_{r=0}^g D_j \mathcal F_r D_k \mathcal F_{g-r} + 
 D_j D_k \mathcal F_{g-1} ) \;,
\mylabel{eqn:BCOVhae}
\end{equation}
where $G^{i\bar j}$ is the inverse of the Weil-Peterson metric 
$G_{i \bar j}=\pd_{x_i}\pd_{\bar x_j} K(x,\bar x)$ and $D_j: 
T^{1,0}\mathcal M^{cpl}(X^\vee)\otimes \mathcal L^{\otimes n} 
\rightarrow T^{1,0}\mathcal M^{cpl}(X^\vee)\otimes \mathcal L^{\otimes n}$ 
is the covariant derivative, which acts on a vector field $Z^k$ taking 
value on $\mathcal L^{\otimes n}$ by $D_j Z^k =\pd_{x_j}Z^k + 
\sum_l \Gamma_{jl}^k Z^l -n {\pd_{x_j} K} Z^k$ where $\Gamma_{jl}^k$ 
is the metric connection. 
As we see here, the holomorphic anomaly equation (\ref{eqn:HAEa}) 
is very similar to BCOV holomorphic anomaly equation. They share 
similar forms, however, associated meaning seems to be slightly 
different. For example, 
in the case of BCOV equation (\ref{eqn:BCOVhae}), 
the holomorphic ambiguity arises 
from the nontrivial holomorphic sections of $\mathcal L^{2-2g}$, which 
we write hereafter  
$f_{g}(x) \in H^0(\mathcal M^{cal}(X^\vee), \mathcal L^{2-2g})$. 
In the end of this subsection, we will compare this holomorphic ambiguity 
with that of $f_{2g+6n-2}(E_4,E_6)$ for (\ref{eqn:HAEa}).

In [BCOV1,2], the general solution of the the holomorphic 
anomaly equation (\ref{eqn:BCOVhae}) has been constructed using the 
K\"ahler geometry (, more precisely {\it special K\"ahler geometry}, ) 
on the moduli space $\mathcal M^{cpl}(X^\vee)$. 
There it was also found that the solutions give the generating functions 
of higher genus Gromov-Witten invariants,
$\mathtt F_g(t)=\sum_\beta  
N_g(\beta)q^\beta \;(q^{2\pi \sqrt{-1} t})$. Namely it is claimed that 
when we introduce  the flat coordinate 
$t_i=t_i(x)$ characterized by $\Gamma_{t_i t_j}^{t_k}=0$ and a property   
$t_i \sim \frac{1}{2\pi \sqrt{-1}} log \, x_i$ near the large complex 
structure limit point, then the generating functions 
will be given by $\mathtt F_g(t):=(w_0(x))^{2g-2} \mathcal F_g(x)$.  
Where $w_0(x)$ is the unique period integral 
which is regular at the large complex structure limit point and 
behaves like $w_0(x)=1+O(x)$ near that point.  General recursive 
formula valid for all genera may be found in [BCOV2], however   
for simplicity, here we only reproduce their results 
for the case of genus two. 

\vskip0.5cm

\noindent
{\it (Solution of BCOV holomorphic anomaly equation at $g=2$)  
Assume the generating functions $\mathtt F_0(t)$ and $\mathtt F_1(t)$ 
are determined. Then there exist propagators 
$\mathtt S^{t_it_j}, \mathtt S^{t_i\phi}, 
\mathtt S^{\phi\phi}$ (symmetric tensors on $\mathcal M^{cal}(X^\vee)$), and 
a holomorphic section $f_{2}(x)$ of $\mathcal L^{\otimes 2}$ which 
express the genus generating function $\mathtt F_2(t)$ by 
$$
\begin{aligned}
\mathtt F_2(t) &= {1\over2}\sum_{j,k} \mathtt S^{t_it_j} 
\left( \pd_{t_j} \pd_{t_k} \mathtt F_1 + 
       \pd_{t_j} \mathtt F_1 \pd_{t_k} \mathtt F_1 \right) \cr
& -{1\over4} \sum_{j,k,m,n} 
  \mathtt S^{t_jt_k}\mathtt S^{t_mt_n}\left( {1\over2} 
\mathtt K_{t_jt_kt_mt_n}+
  2 \mathtt K_{t_mt_nt_j} \pd_{t_k} \mathtt F_1\right) 
 + {\chi \over 24} \sum_{k} \mathtt S^{t_k\phi} \pd_{t_k} \mathtt F_1 \cr
\end{aligned}
$$
\begin{equation}
\begin{aligned}
& +{1\over8} \sum_{j,k,m,n,r,s} \mathtt S^{t_jt_k} 
\mathtt S^{t_mt_n}\mathtt S^{t_rt_s} 
    \mathtt K_{t_mt_nt_j}\mathtt K_{t_kt_rt_s} \\
&+{1\over12}\sum_{a,b,j,k,m,n} \mathtt K_{t_at_jt_m}\mathtt K_{t_bt_kt_n}
   \mathtt S^{t_at_b}\mathtt S^{t_jt_k}\mathtt S^{t_mt_n} \hskip3cm \; \\
& -{\chi \over 48} \mathtt S^{t_j\phi}
  \mathtt S^{t_mt_n} \mathtt K_{t_mt_nt_j} 
  +{\chi \over24}({\chi  \over 24}-1) \mathtt S^{\phi\phi}  
  + w_0(x)^{2}f_2(t) \;, \\
\end{aligned} 
\end{equation}
where $\chi=\chi(X)$ is the Euler number of $X$ and 
$\phi$ is called dilaton. Also we set, three point and four point functions, 
respectively, by  
$$
\mathtt K_{t_mt_nt_j} := \pd_{t_m}\pd_{t_n} \pd_{t_j} \mathtt F_0(t) \;,\; 
\mathtt K_{t_mt_nt_jt_k} := \pd_{t_m}\pd_{t_n} \pd_{t_j} \pd_{t_k}
\mathtt F_0(t) \;.\; 
$$  
}

\noindent
{\bf Remark.} (1) In the above formula, the propagator $\mathtt S^{t_it_j}$ 
is determined by solving relation 
$\sum_m \mathtt K_{t_it_jt_m}\mathtt S^{t_mt_k}= 
\pd_{t_i}K \delta_{t_j}^{t_k}+\pd_{t_j}K 
\delta_{t_i}^{t_k} - \Gamma_{t_i t_j}^{t_k}$, which arises from 
special K\"ahler geometry on $\mathcal M^{cpl}(X^\vee)$. Other propagators 
$\mathtt S^{t_i\phi}$ and $\mathtt S^{\phi\phi}$ are also determined 
by similar relations. 
Determining these propagators is one of the most difficult parts 
to construct the solutions. Once these are determined, 
$\mathtt F_g(t) (g\geq 2)$ are 
determined summing over several terms which are in 1 to 1 
corresponding to the graphs representing degenerations of genus $g$ 
curves (see [BCOV1,2]).  For each genus, we have to fix the holomorphic 
ambiguity $f_g(x)$ by vanishing conditions like those discussed in section 
3.1. 

\noindent
(2) The flat coordinate $t_i=t_i(x)$ is called {\it mirror map}. It relates 
the complex structure moduli space of $X^\vee$ to the complexified K\"ahler 
cone $H^2(X,\bold R)+\sqrt{-1}\mathcal K_X$. Then by the coordinate 
$(t_1,\cdots,t_r)$, we understand a point $\sum_i t_k J_k \in 
H^2(X,\bold R)+\sqrt{-1}\mathcal K_X$ with some positive integral 
generators $J_1,\cdots,J_r$ of $H^2(X,\bold Z)$. See e.g. [HLY] for details. 
When some of the integral generators, say $J_r$, represents Poincar\'e 
dual of a smooth divisor $S$ (with $K_S>0$), then the limit $Im(t_r) 
\rightarrow \infty$ is called {\it local mirror symmetry limit}. 
Projective space $\bold P^2$,  del Pezzo surfaces (and also 
rational elliptic surfaces) as smooth divisor in Calabi-Yau threefolds 
are well-studied examples (see [CKYZ]).

\vskip0.5cm

As remarked above, constructing solutions of BCOV holomorphic anomaly 
equation involves three steps; 1) finding the propagators, 2) summing 
over graphs parametrizing the degeneration, 3) fixing the holomorphic 
ambiguity. Since all of them are technically so involved that it is very 
hard to make solutions $\mathtt F_g$ in general. However it has been 
found in [KZ] that under {\it local mirror symmetry limit} the solutions for 
$\mathtt F_g$ are considerably simplified. 

\vskip0.2cm
\noindent
{\it (Local mirror symmetry limit [KZ]) Under the local 
mirror symmetry limit to a smooth divisor, if it exists, 
the both propagators $\mathtt S^{t_i\phi}$ and $\mathtt S^{\phi\phi}$ vanish. 
In other words, the dilaton $\phi$ does not propagate. 
}

\vskip0.2cm 
As we see in the genus two example (\refFtwoBCOV), the local limit 
simplifies the form of $\mathtt F_g$. However its manipulation is still 
tedious unless $\mathtt S^{t_it_j}=S_i \delta^{t_it_j}$.  For the local mirror 
symmetry limit to a rational elliptic surface $S$, our observation 
is that this simplification is in fact the case!

\subsection{$\mathbf{S=\frac{1}{2}K3}$} Here we present the form of the 
propagator $\mathtt S^{t_it_j}$ for rational elliptic surfaces, i.e. 
$S=\frac{1}{2}K3$. The main observation is the compatibility of the 
holomorphic anomaly equation (\ref{eqn:HAEa}) studied in detail 
in section 3 with the recursion relation (\refFtwoBCOV) 
which follows from  BCOV holomorphic anomaly equation. 

\begin{definition} \thlabel{th:FgZgn} Let $Z_{g;n}(q) \; 
(q=e^{2\pi \sqrt{-1} \tau})$ be the solutions of the holomorphic anomaly 
equation (\ref{eqn:HAEa}). Then we define a series 
\begin{equation}
\mathtt F_g(q,p)  := \sum_{n\geq 1} Z_{g;n}p^n \;\;.
\mylabel{eqn:FgDef}
\end{equation}
\end{definition}

Now let us introduce the following hypergeometric series: 
\begin{align*}
&w_0(x,y):=\sum_{n,m\geq0} c(n,m)x^ny^m  \\
c(n,m):=&\frac{\Gamma(1+6n)}{\Gamma(1+3n)\Gamma(1+2n)\Gamma(1+n-m)
\Gamma(1+m)^2\Gamma(1-m)} 
\end{align*}
The mirror map or the flat coordinate $t_1,t_2$ may be defined by 
this hypergeometric series; 
$$
t_i:=\frac{1}{2\pi \sqrt{-1}} 
    \frac{\pd_{\rho_i} w_0(x,y,\rho_1,\rho_2)}{w_0(x,y)} \vert_{\rho_i=0}\;,
$$
where $w_0(x,y,\rho_1,\rho_2):=\sum_{n,m\geq0}c(n+\rho_1,m+\rho_2)x^{n+\rho_1}
y^{m+\rho_2}$. We denote the inverse relation of $t_i=t_i(x,y) \; (i=1,2)$ 
by $x=x(q,p), y=y(q,p)$ setting $q=e^{2\pi \sqrt{-1} t_1}, 
p=e^{2\pi \sqrt{-1} t_2}$. Detailed analysis of the mirror map can 
be found in [HST1], and following the method there it is straightforward 
to see $x=x(q)$ and $y=y(q,p)$, i.e. the relations are lower triangular. 
Furthermore it is easy to derive 
\begin{equation}
x(q)(1-432 x(q))=\frac{1}{j(q)} \;\;,\;\; 
w_0(x(q),y(q,p))=w_0(x(q)) = E_4(q)^{\frac{1}{4}} \;\;,
\mylabel{eqn:mirrormapx}
\end{equation}
where $j(q)$ is the elliptic modular function and $w_0(x,y)=w_0(x)$ from 
the definition. The next statement 
follows directly from the derivation of the holomorphic anomaly 
equation given in [HST1], changing the parametrization there in an obvious 
way. 

\begin{proposition} 
The functions $\mathtt F_0(q,p)$ and $\mathtt F_1(q,p)$ 
defined above may be written 
by the hypergeometric series $w_0(x,y), \pd_{\rho_1}w_0(x,y), \pd_{\rho_2} 
w_0(x,y)$ and $\pd_{\rho_1}\pd_{\rho_2}w_0(x,y)$. 
Especially $\mathtt F_1(q,p)$ is given by 
\begin{equation}
\mathtt F_1(q,p)=\frac{1}{2}\log\Big\{ [ (1-432x(1-y))(1-y) ]^{-\frac{1}{6}} 
\frac{\pd y}{\pd t_2} \Big\}\Big|_{x=x(q),y=y(q,p)}\;\;,
\mylabel{eqn:Fonexy}
\end{equation}
where $(1-432x(1-y))(1-y)=:dis_0$ is a component 
of the discriminant, which follows from the characteristic variety 
of the differential equation satisfied by the hypergeometric 
series.   
\end{proposition}

{\bf Remark.} (1) The discriminant from the differential 
equation may be found to be  $x y (1-432x)^3 dis_0$, 
where the normal crossing divisors $x=0$ and $y=0$ give rise to 
the large complex structure limit.  

\noindent
(2) As is evident from the context, the flat coordinate $t_1$ should 
be identified with the modular parameter $\tau$ in (\ref{eqn:HAEa}). 
Then the holomorphic anomaly ( or modular anomaly) in (\ref{eqn:HAEa}) 
comes from the `anomalous' modular transformation;
\begin{equation*}
E_2(\tau)\big\vert_{\tau \rightarrow \frac{a\tau+b}{c\tau+d}} 
=(c\tau + d)^2 E_2(\tau)+\frac{12}{2\pi \sqrt{-1}} c(c\tau+d), 
\end{equation*}
where $\left(\begin{matrix} a & b \\ c & d \\ \end{matrix}\right)$ 
is an element in  $PSL(2,\bold Z)$. 
As we have $x=x(q)=\frac{1\pm \sqrt{1-1728/j(q)}}{864}$ which is 
modular function (for a modular subgroup of index two), the modular 
anomaly should be traced to the form $y=y(q,p)$. Following exactly the same 
calculations presented in [HST1], we can determine $E_2(q)$-dependence 
of $y(q,p)$, and from which we can derive 
\begin{equation}
y(q,p) \big\vert_{t_1 \rightarrow \frac{at_1+b}{ct_1+d}} 
= y(q,p) e^{-\frac{1}{2\pi \sqrt{-1}} c(ct_1+d) \pd_{t_2} \mathtt F_0(q,p)}
\mylabel{eqn:anomalyY} \;\;.
\end{equation}
Using this relation essentially, we can prove that  
$\mathtt F_1(q,p)$ given in (\ref{eqn:Fonexy}) in fact satisfies 
the holomorphic anomaly equation (\ref{eqn:HAEa}) with $g=1$.

\vskip0.3cm

Now for our higher genus function $\mathtt F_g(q,p) \; (g \geq 2)$, we may 
observe the following:

\vskip0.3cm

\begin{conjecture} 
Define the propagator $\mathtt S^{t_it_j}$ by 
$\mathtt S^{t_1t_1}=\mathtt S^{t_1t_2}=\mathtt S^{t_2t_1}=0$ and 
\begin{equation*}
\mathtt S^{t_2t_2}=-{1\over \mathtt K_{t_2t_2t_2}}\frac{\pd \;}{\pd t_2}
\log \left( y \frac{\pd t_2}{\pd y} \right) \;\;, 
\end{equation*}
and also 
$ \mathtt K_{t_2t_2t_2}:=\pd_{t_2}\pd_{t_2}\pd_{t_2} \mathtt F_0(q,p)$. 
Then there exists a rational function $f_g(x,y)$ of the form 
$$
f_g(x,y)=(\text{polynomial in x, y})/(dis_0)^{2g-2},
$$
which reproduces our function $\mathtt F_g(q,p)$ in (\ref{eqn:FgDef}) 
from the BCOV recursion relation with vanishing dilaton propagators 
(e.g. the recursion formula (\refFtwoBCOV) for $g=2$ with 
$\mathtt S^{t_i\phi}=\mathtt S^{\phi\phi}=0$. )
\end{conjecture}

\vskip0.3cm

For example, for $g=2$ we have the reduced BCOV recursion relation,
\begin{align*}
\mathtt F_2(q,p)=& 
  {1\over2} \mathtt S^{t_2t_2} 
\left( \pd_{t_2} \pd_{t_2} \mathtt F_1 + 
       \pd_{t_2} \mathtt F_1 \pd_{t_2} \mathtt F_1 \right) 
 -{1\over8} 
  \mathtt S^{t_2t_2}\mathtt S^{t_2t_2}\left( \mathtt K_{t_2t_2t_2t_2}+
  4 \mathtt K_{t_2t_2t_2} \pd_{t_2} \mathtt F_1\right)  \\
& +{5\over24}  \mathtt S^{t_2t_2} \mathtt S^{t_2t_2}\mathtt S^{t_2t_2} 
    \mathtt K_{t_2t_2t_2}\mathtt K_{t_2t_2t_2}  + w_0^{2}(q)f_2(q,p) \;,  
\mylabel{eqn:Ftwolocal}
\end{align*}
with the holomorphic ambiguity $f_2(x,y)$. We may verify directly that 
our functions $\mathtt F_0, \mathtt F_1$ and $\mathtt F_2$ satisfy 
the above recursion relation with 
\begin{align*}
&f_2(x,y)= 1\Big/\Big( 240 (1 - 432 x (1 - y))^2  (1 - y)^2 \Big) \times  \\
&\Big(
   \left( 1 - 72 x - 311040 {x^2} + 67184640 {x^3} \right) y     
   - 1430  \left(x - 1296 x^2 + 373248 x^3 \right) {y^2} \\
 & + 2\left (751x - 1386720 x^2 + 599063040 x^3 \right){y^3} 
  + 1231200 \left({x^2} - 864{x^3} \right) {y^4}  \\
 & + 332190720{x^3}{y^5} \Big) 
\end{align*}
For $g=3$, the corresponding recursion relation for $\mathtt F_3(q,p)$ 
follows directly from [BCOV2] (, see also [KKV]). 
We can also find the rational function $f_3(x,y)$ 
of the form stated above, although we do not reproduce its lengthy 
form here.

{\bf Remark.} (1) $f_g(x,y)$ is the holomorphic ambiguity in the 
solutions of BCOV holomorphic anomaly equation. As clear from 
(\ref{eqn:mirrormapx}) 
and (\ref{eqn:anomalyY}), 
$w_0(x)^{2g-2} f_g(q,p)$ does not behave as a modular form under 
$t_1 \rightarrow (at_1+b)/(ct_1+d)$.  This means that the holomorphic 
ambiguity in the solutions of BCOV equation differs from the ambiguity 
$\sum_n f_{2g+6n-2}(E_4(q),E_6(q)) p^n$ which arises when solving 
the holomorphic anomaly equation (\ref{eqn:HAEa}). 

\noindent 
(2) It is worth while writing here the form of the propagator 
$\mathtt S^{t_it_j}$ in the coordinate $x,y$, i.e. 
that defined by $\mathtt S^{t_it_j}=w_0(x)^2 
\sum_{k,l} \mathtt S^{x_kx_l} \frac{\pd t_i}{\pd x_k} 
\frac{\pd t_j}{\pd x_l}$. 
After some calculation, it is easy to derive 
$\mathtt S^{xx}=\mathtt S^{xy}=0$ and  
$$
\mathtt S^{yy}=
{1\over \mathtt K_{yyy}}\left( -\Gamma_{y\;y}^{\;y}-{1\over y} \right) 
\;\;, \;\;
\Gamma_{y\;y}^{\;y}=
{\pd y \over \pd t_2} {\pd \; \over \pd y}
\left({\pd t_2 \over \pd y}\right) \;\;,
$$
where $\mathtt S^{t_2t_2}=w_0^2 \; \mathtt S^{yy} 
({\pd t_2 \over \pd y})^2$ and $\mathtt K_{yyy}
=w_0(x)^2 (\frac{\pd t_2}{\pd y})^3 \mathtt K_{t_2t_2t_2}$. Note 
that we have $\frac{\pd y}{\pd t_2}\frac{\pd t_2}{\pd y}=1$ since 
$x=x(q), y=y(q,p)$ with $q=e^{2\pi \sqrt{-1}t_1}, p=e^{2\pi \sqrt{-1} t_2}$.

\subsection{ $\bold P^1 \times \bold P^1$ }

As a slightly different two parameter model, we may consider 
a local limit to a smooth divisor $\bold P^1 \times \bold P^1$ in 
a Calabi-Yau 3-fold. A Calabi-Yau model containing this surface 
may be realized as an elliptic fibration over $\bold P^1 \times \bold P^1$.  
The local mirror limit is a limit in which the volume of the 
fiber goes to infinity. And the resulting space may be identified as 
a non-compact Calabi-Yau manifold, $K_{\bold P^1 \times \bold P^1} 
\rightarrow \bold P^1 \times \bold P^1$.  Then the cohomology classes 
of compact support my be identified with those of the base space 
$\bold P^1 \times \bold P^1$. For a positive basis of 
$H^2(\bold P^1 \times \bold P^1,\bold Z)$, we choose the hyperplane classes 
$H_1$ and $H_2$ from each $\bold P^1$. Then under the local mirror symmetry 
limit, we have the generating function for Gromov-Witten invariants of 
$\bold P^1 \times \bold P^1$ which we parametrize by  
$$
\mathtt{F}_g(q,p)=\sum_{\beta \in H^2(\bold P^1\times\bold P^1,\bold Z)} 
N_g(\beta) q^{(\beta,H_2)} p^{(\beta,H_1+H_2)} \;\;.
$$
Where a special parametrization for $q,p$ has been chosen  so that 
we can utilize the Segre embedding, $\bold P^1 \times \bold P^1$ into 
$\bold P^3$ as degree 2 surface. Namely, the diagonal direction $H_1+H_2$ 
may be identified with the class coming from the hyperplane class of 
$\bold P^3$. The reduction of BCOV holomorphic anomaly equation to 
the diagonal one parameter subspace ($q=1$) has been studied in [KKV], and 
our parametrization naturally recovers two parameters, $H_2$ and $H_1+H_2$, 
from this one parameter reduction. Since the calculations are parallel to 
those appeared in $\frac{1}{2}K3$ case, here we simply write corresponding 
formulas for $\mathtt F_g(q,p)$.

The hypergeometric series we start with is given by
\begin{footnote}{
This corresponds to the choice of the ``charge vector'' 
$
l^{(1)}=( 0;0,0,-1,-1,1,1, 0), 
l^{(2)}=( 0;0,0, 1, 1,0,0,-2), 
l^{(3)}=(-6;3,2, 0, 0,0,0, 1)$ 
for the elliptic Calabi-Yau threefold.}
\end{footnote}
\begin{equation*}
w_0(x,y)=\sum_{n,m \geq 0} c(n,m)x^ny^m , \;\;
c(n,m)=\frac{1}{\Gamma(1-n+m)^2\Gamma(1+n)^2\Gamma(1-2m)}. 
\end{equation*}
As before, the mirror map is defined by 
$2\pi \sqrt{-1} t_i= 
\frac{\pd_{\rho_i} w_0(x,y,\rho_1,\rho_2)}{w_0(x,y)}|_{\rho_i=0}$. Then 
again we find a lower triangular form for $x=x(q,p), y=y(q,p)$ as 
\begin{align*}
&x = q \cr
&y= p -(2 + 2 q ){p^2} + ( 3 + 3 {q^2} ){p^3} - 
    ( 4 + 4 q + 4 {q^2} + 4{q^3} ) {p^4} + O(p^5) \;.
\end{align*}
By using mirror symmetry, we can write $\mathtt F_0(q,p)$ 
in terms of hypergeometric series $w_0(x,y)=1, \pd_{\rho_1}w_0(x,y), 
\pd_{\rho_2}w_0(x,y)$ and $\pd_{\rho_1}\pd_{\rho_2}w_0(x,y)$. 
The genus one function and the propagator has similar form as before;
\begin{align*}
\mathtt F_1(q,p)
&= \frac{1}{2}
   \log \left\{ \left(1+16y^2(1-x)^2-8y(1+x) \right)^{-{1\over6}} 
            y^{-{7\over6}} 
           {\pd y \over \pd t_2} \right\} \;\;,  \\ 
\mathtt S^{t_2t_2}&=-{1\over \mathtt K_{t_2t_2t_2}}{\pd \; \over \pd t_2}
\log \left( y \frac{\pd t_2}{\pd y} \right) \;\;, \\
\end{align*}
with $\frac{1}{(2\pi \sqrt{-1})^3}\mathtt K_{t_2t_2t_2}=
\frac{1}{(2\pi \sqrt{-1})^3} \pd_{t_2}\pd_{t_2}\pd_{t_2} \mathtt F_0(q,p)=
-1-(2 + 2 q)p - (2 + 32 q + 2 q^2 )p^2 + \cdots$. 
When we write the propagator in $x,y$ coordinate we have 
$$
\mathtt S^{yy}={1\over \mathtt K_{yyy}}
\left( -\Gamma_{y\;y}^{\;y}-{1\over y} \right) 
\;\;, \;\;
\Gamma_{y\;y}^{\;y}=
{\pd y \over \pd t_2} {\pd \; \over \pd y}
\left({\pd t_2 \over \pd y}\right) \;\;,
$$
where $\mathtt S^{t_2t_2}=w_0^2 \; \mathtt S^{yy} 
({\pd t_2 \over \pd y})^2$ and $\mathtt K_{yyy}
=w_0(x)^2 (\frac{\pd t_2}{\pd y})^3 \mathtt K_{t_2t_2t_2}$.

Now  BCOV recursion formula for $\mathtt F_2$ is the same as  
the previous case (\refFtwoBCOV), and for the holomorphic ambiguity 
$f_2(x,y)$ we find  
\begin{align*}
f_2 = &\Big(-11 y (1+x) 
       + 12 y^2 (31+58x+31x^2)  
       - 16 y^3 (333+595 x+595x^2+333x^3)  \\   
     &   + 64 y^4 (1-x)^2 (551+994 x + 551 x^2)   
      - 107520 y^5 (1-x)^4 (1+x)             \\
     & + 122880 y^6 (1-x)^6  \Big) \Big/ 
        \left( 720 \left(1 -8 y (1+x) + 16 y^2 (1-x)^2 \right)^2 \right) \;\;.
\end{align*}
Here the holomorphic ambiguity $f_2(x,y)$ has been fixed by requiring 
the vanishing for BPS numbers $n_2(a H_1+b H_2)$ for lower degrees $a, b$, 
and one known result $n(4 H_1 + 2 H_2)=116$ in [KKV]. 

In the following tables, we have listed the BPS numbers 
$n_g(a,b)=n_g(a H_1+ b H_2)$ up to genus two, which result from 
the Gopakumar-Vafa formula (\ref{eqn:GVformula}).

$$
\vbox{\offinterlineskip
\hrule
\halign{ \strut 
\vrule#  
& $\;$ \hfil #  \hfil   
&&\vrule#  
& $\;$ \hfil # \hfil 
& \hfil # \hfil  
& \hfil # \hfil  
& \hfil # \hfil 
& \hfil # \hfil  
& \hfil # \hfil  
& \hfil # \hfil  
& \hfil # \hfil  
\cr 
&a $\backslash$ b && 0 & 1 & 2 & 3 & 4 & 5 & 6 & 7 &\cr
\noalign{\hrule} 
&0 &&  0 & -2 & 0  &  0 &   0 &   0 &   0 &  0&\cr
&1 && -2 & -4 & -6 & -8 & -10 & -12 & -14 & -16 &\cr
&2 &&  0 & -6 & -32 & -110 & -288 & -644 & -1280 & -2340 &\cr
&3 &&  0 & -8 & -110 & -756 & -3556 & -13072&-40338 & -109120&\cr 
&4 &&  0 & -10 & -288 & -3556 & -27264 & -153324 & -690400 &-2627482 &\cr
&5 &&  0 & -12 & -644 & -13072 &-153324  & -1252040 & -7877210 & -40635264&\cr
&6 &&  0 & -14 & -1280 & -40338 & -690400 & -7877210 &-67008672 & -455426686
       &\cr
&7 &&  0 & -16 & -2340 & -109120 & -2627482 & -40635264 &  -455426686
        &-3986927140 &\cr
}
\hrule} 
$$
{\leftskip1cm\rightskip1cm\noindent
{\bf Table 6.} Genus zero BPS numbers  
$n_0(a,b)=n_0(a H_1 + b H_2)$.    \par} 

$$
\vbox{\offinterlineskip
\hrule
\halign{ \strut 
\vrule#  
& $\;$ \hfil #  \hfil   
&&\vrule#  
& $\;$ \hfil # \hfil 
& \hfil # \hfil  
& \hfil # \hfil  
& \hfil # \hfil 
& \hfil # \hfil  
& \hfil # \hfil  
& \hfil # \hfil  
& \hfil # \hfil  
\cr 
&a $\backslash$ b && 0 & 1 & 2 & 3 & 4 & 5 & 6 & 7 &\cr
\noalign{\hrule} 
&0 &&  0 &  0 & 0  &  0 &   0 &   0 &   0 &  0&\cr
&1 &&  0 &  0 & 0  &  0 &   0 &   0 &   0 &  0 &\cr
&2 &&  0 &  0 & 9 &  68& 300 & 988 & 2698 &  6444&\cr
&3 &&  0 &  0 & 68 & 1016 & 7792 &41376 & 172124& 599856&\cr 
&4 &&  0 &  0 & 300 & 7792 & 95313 & 760764 & 4552692 & 22056772 &\cr
&5 &&  0 &  0 & 988 & 41376 & 760764  & 8695048 & 71859628 & 467274816&\cr
&6 &&  0 &  0 & 2698 & 172124 & 4552692 & 71859628 &795165949 & 6755756732
       &\cr
&7 &&  0 &  0 & 6444 & 599856 & 22056772 & 467274816 & 6755756732
        & 73400088512 &\cr
}
\hrule} 
$$
{\leftskip1cm\rightskip1cm\noindent
{\bf Table 7.} Genus one BPS numbers  
$n_1(a,b)=n_1(a H_1 + b H_2)$.    \par} 

$$
\vbox{\offinterlineskip
\hrule
\halign{ \strut 
\vrule#  
& $\;$ \hfil #  \hfil   
&&\vrule#  
& $\;$ \hfil # \hfil 
& \hfil # \hfil  
& \hfil # \hfil  
& \hfil # \hfil 
& \hfil # \hfil  
& \hfil # \hfil  
& \hfil # \hfil  
& \hfil # \hfil  
\cr 
&a $\backslash$ b && 0 & 1 & 2 & 3 & 4 & 5 & 6 & 7 &\cr
\noalign{\hrule} 
&0 && 0&0 & 0 & 0 & 0 & 0 & 0 & 0 &\cr
&1 && 0&0 & 0 & 0 & 0 & 0 & 0 & 0 &\cr
&2 && 0&0 & 0 & -12 & -116 & -628 & -2488 & -8036 &\cr
&3 && 0&0 & -12 & -580 & -8042 & -64624 & -371980 & -1697704&\cr 
&4 && 0&0 & -116 & -8042 & -167936 & -1964440 & -15913228 & -99308018&\cr
&5 && 0&0 & -628 & -64624 & -1964440 & -32242268 & -355307838 & -2940850912&\cr
&6 && 0&0 & -2488 & -371980& -15913228& -355307838& -5182075136& -55512436778
       &\cr
&7 && 0&0 & -8036 & -1697704 & -99308018 & -2940850912 & -55512436778 
        & -754509553664&\cr
}
\hrule} 
$$
{\leftskip1cm\rightskip1cm\noindent
{\bf Table 8.} Genus two BPS numbers  
$n_2(a,b)=n_2(a H_1 + b H_2)$.  \par} 




\end{document}